\documentclass[useAMS,usenatbib]{mn2e}

\usepackage{dsfont}
\usepackage{amsmath}
\usepackage{latexsym,times}

\usepackage{amssymb}
\usepackage{graphicx}
\usepackage{verbatim}
\usepackage{natbib}
\usepackage{eucal}
\usepackage{calligra}
\usepackage[pagewise]{lineno}

\usepackage{pifont}

\usepackage{longtable}
\usepackage{multicol}
\usepackage{booktabs}

\usepackage[normalem]{ulem}

\usepackage[usenames,dvipsnames]{color}

\DeclareMathAlphabet{\mathscr}{OT1}{pzc}%
                                 {m}{it}
                                 
\newcommand{\mnras}{MNRAS}
\newcommand{\jcap}{JCAP}
\newcommand{\apj}{ApJ}
\newcommand{\apjs}{ApJS}
\newcommand{\aj}{AJ}
\newcommand{\aap}{A\&A}
\newcommand{\prd}{Phys. Rev. D}

\newcommand{\physrep}{PhR}
\newcommand{\procspie}{SPIE}
\newcommand{\aaps}{A\&AS}
\def\physrep{Phys.~Rep.}   
\def\apjl{ApJL}

\newcommand{\be}{\begin{equation}}
\newcommand{\ee}{\end{equation}}
\newcommand{\bes}{\begin{equation*}}
\newcommand{\ees}{\end{equation*}}
\newcommand{\bea}{\begin{eqnarray}}
\newcommand{\eea}{\end{eqnarray}}
\newcommand{\beas}{\begin{eqnarray*}}
\newcommand{\eeas}{\end{eqnarray*}}

\newcommand{\matr}[1]{\mathbf{#1}}

\newcommand{\mpch}{\;{\rm Mpc}/h}
\newcommand{\de}{{\rm d}}

\newcommand{\zl}{z_{\rm L}}
\newcommand{\zs}{z_{\rm s}}

\newcommand{\mr}{\mathrm}
\newcommand{\ensav}[1]{\left\langle #1 \right\rangle}

\def\ave#1{\left\langle #1 \right\rangle}
\newcommand{\SE}{single-epoch}
\newcommand{\ME}{multi-epoch}

\newcommand{\iband}{$i$-band}
\newcommand{\zband}{$z$-band}

\newcommand{\grizY}{$g,r,i,z,Y$}

\newcommand{\photoz}{photo-$z$}

\newcommand{\nomepochs}{10}
\newcommand{\nomdepth}{24.1}
\newcommand{\ngmix}{\texttt{ngmix}}
\def\redmagic{redMaGiC}

\begin{document}

%\linenumbers

\title[DES Galaxy-Galaxy Lensing]
  {Galaxy-Galaxy Lensing in the DES Science Verification Data}
\author[Clampitt et al.]
{\parbox{\textwidth}{
J.~Clampitt$^1$,\thanks{E-Mail: clampitt@sas.upenn.edu}
C.~S{\'a}nchez$^2$,
J.~Kwan$^1$,
E.~Krause$^3$,
N.~MacCrann$^4$,
Y.~Park$^{5,6}$,
M.~A.~Troxel$^4$,
B.~Jain$^1$,
E.~Rozo$^7$,
E.~S.~Rykoff$^{3,8}$,
R.~H.~Wechsler$^{9,3,8}$,
J.~Blazek$^{10}$,
C.~Bonnett$^2$,
M.~Crocce$^{11}$,
Y.~Fang$^1$,
E.~Gaztanaga$^{11}$,
D.~Gruen$^{3,8,\dagger}$,
M.~Jarvis$^1$,
R.~Miquel$^{12,2}$
J.~Prat$^2$,
A.~J.~Ross$^{10}$,
E.~Sheldon$^{13}$,
J.~Zuntz$^4$,
T. M. C.~Abbott$^{14}$,
F.~B.~Abdalla$^{15,16}$,
R.~Armstrong$^{17}$,
M.~R.~Becker$^{9,3}$,
A.~Benoit-L{\'e}vy$^{18,15,19}$,
G.~M.~Bernstein$^1$,
E.~Bertin$^{18,19}$,
D.~Brooks$^{15}$,
D.~L.~Burke$^{3,8}$,
A. Carnero Rosell$^{20,21}$,
M.~Carrasco~Kind$^{22,23}$,
C.~E.~Cunha$^{3}$,
C.~B.~D'Andrea$^{24,25}$,
L.~N.~da Costa$^{20,21}$,
S.~Desai$^{26,27}$,
H.~T.~Diehl$^{28}$,
J.~P.~Dietrich$^{26,27}$,
P.~Doel$^{15}$,
J.~Estrada$^{28}$,
A.~E.~Evrard$^{29,30}$,
A.~Fausti Neto$^{20}$,
B.~Flaugher$^{28}$,
P.~Fosalba$^{11}$,
J.~Frieman$^{28,6}$,
R.~A.~Gruendl$^{22,23}$,
K.~Honscheid$^{10,31}$,
D.~J.~James$^{14}$,
K.~Kuehn$^{32}$,
N.~Kuropatkin$^{28}$,
O.~Lahav$^{15}$,
M.~Lima$^{32,33}$,
M.~March$^{1}$,
J.~L.~Marshall$^{34}$,
P.~Martini$^{10,35}$,
P.~Melchior$^{17}$,
J.~J.~Mohr$^{26,27,36}$,
R.~C.~Nichol$^{24}$,
B.~Nord$^{28}$,
A.~A.~Plazas$^{37}$,
A.~K.~Romer$^{38}$,
E.~Sanchez$^{39}$,
V.~Scarpine$^{28}$,
M.~Schubnell$^{30}$,
I.~Sevilla-Noarbe$^{39}$,
R.~C.~Smith$^{14}$,
M.~Soares-Santos$^{28}$,
F.~Sobreira$^{20}$,
E.~Suchyta$^{40}$,
M.~E.~C.~Swanson$^{23}$,
G.~Tarle$^{30}$,
D.~Thomas$^{24}$,
V.~Vikram$^{41}$,
A.~R.~Walker$^{14}$}
  \vspace{0.4cm}\\
\parbox{\textwidth}{Author affiliations are listed at the end of this paper\\ $\star$ E-mail: \texttt{\rm \texttt{clampitt@sas.upenn.edu}}}}

\maketitle

\begin{abstract}
We present galaxy-galaxy lensing results from 139 square degrees of Dark Energy Survey (DES) Science Verification (SV) data.
Our lens sample consists of red galaxies, known as redMaGiC, which are specifically selected to have a low photometric redshift error and outlier rate.
The lensing measurement has a total signal-to-noise of 29 over scales $0.09 < R < 15$
Mpc/$h$, including all lenses over a wide redshift range $0.2 < z < 0.8$.
Dividing the lenses into three redshift bins for this constant moving number density sample, we find no evidence for evolution in the halo mass with redshift.
We obtain consistent results for the lensing measurement with two independent shear pipelines, \texttt{ngmix} and \texttt{im3shape}.
We perform a number of null tests on the shear and photometric redshift catalogs and quantify resulting systematic uncertainties.
Covariances from jackknife subsamples of the data are validated with a suite of 50 mock surveys.
The results and systematics checks in this work provide a critical input for future cosmological and galaxy evolution studies with the DES data and redMaGiC galaxy samples.
We fit a Halo Occupation Distribution (HOD) model, and demonstrate that our data constrains the mean halo mass of the lens galaxies, despite strong degeneracies between individual HOD parameters.
\end{abstract}


%%%%%%%%%%%%%%%
\section{Introduction}
%%%%%%%%%%%%%%%

Weak gravitational lensing refers to the subtle distortions in the images of distant galaxies by intervening mass along the line of sight. The measurement of lensing around foreground (lens) galaxies is referred to as galaxy-galaxy lensing \citep{tvj1984, brainerd96, dellantonio96}. Background (source) galaxies are binned in annuli on the sky centered on lens galaxies; the shapes of the background galaxies are projected along the tangential direction and averaged over a population of lens galaxies. The measurement as a function of angular separation can be converted into an estimate of the projected mass profile of the dark matter halos where the galaxies reside.

Galaxy-galaxy lensing measurements have been used to infer the mass distribution within the halos of massive galaxies, the relation of mass to light, the shapes of the halos, and the large-scale galaxy-mass cross-correlation \citep{sjf2004,manetal06,manetal08, cvm2009}. The measurements have many applications, ranging from fitting Navarro Frenk White (NFW) halo mass profiles \citep{nfw1997} to estimating the large-scale bias of galaxies and obtaining cosmological constraints \citep{cvm2009, msb2013, more15}. 
Recent surveys such as CFHTLenS \citep{heymans12, erben13} have presented measurements on galaxy-galaxy lensing \citep{gillis13, vvh14, hudson15}.
Similarly, measurements from KiDS \citep{dejong13, kuijken15} have also studied the galaxy-mass connection using galaxy-galaxy lensing \citep{sifon15, viola15, vanuitert16}.
The galaxy-mass connection has also been studied at high redshift by \citet{ltb12}.

In this paper we measure galaxy-galaxy lensing from Dark Energy Survey (DES) pre-survey Science Verification (SV) data.
DES is an ongoing wide-field multi-band imaging survey that will cover nearly 5000 square degrees of the Southern sky over five years.
With this pre-survey SV data, our goals are to establish the feasibility of measuring galaxy-galaxy lensing with DES, test our measurement pipelines, and make an estimate of the halo properties for a selected galaxy sample.
The detailed tests presented serve as a necessary foundation for other work relying on galaxy-galaxy lensing measurements with these data.
In particular, \citet{kwan17} obtains constraints on cosmological parameters using the combination of galaxy-galaxy lensing and galaxy clustering information with the same data used in this work.
\citet{baxter16} presents complementary cosmological and systematic constraints using the combination of galaxy-galaxy lensing and CMB lensing.
\citet{prat16} measures galaxy-galaxy lensing around a magnitude-limited sample of DES-SV galaxies in order to measure their large-scale bias.
Finally, tangential shear measurements of underdensities such as troughs \citep{gfa15} and voids \citep{sanchez16} also benefit from the tests in this work.

The plan of the paper is as follows.
Section 2 summarizes the basic theory of galaxy-mass correlations and our Halo Occupation Distribution (HOD) model.
Section 3 describes our data: including basic details of DES, descriptions of the lens galaxy sample, pipelines for source galaxy shape measurements, and the photometric redshift estimation of lens and source galaxies.
Our estimators and measurement methodology are presented in Section 4.
Results of null tests that establish the suitability of the shear and photo-$z$ catalogs for galaxy-galaxy lensing are presented in Section 5.
Our measurement results and HOD model fits are in Section 6, as well as discussion of related results in the literature.
We conclude in Section 7.

%%%%%%%%%%%%%%%
\section{Weak-lensing theory and the halo model}
\label{sec:theory}
%%%%%%%%%%%%%%%

Galaxy-galaxy lensing involves the distortion of background galaxy images in the presence of foreground dark matter halos, which are occupied by the lens galaxies.
This distortion makes the background galaxy image stretch tangentially to the line joining the background and foreground galaxies.
The magnitude of this tangential shear, $\gamma_\mr{t}(\theta)$, and of the related excess surface density, $\Delta\Sigma(R)$, provides a means of learning about the local dark matter profile and galaxy environment.

We relate the properties of lens galaxies to the underlying dark matter distribution through Halo Occupation Distribution (HOD) modeling \citep{zheng2005, zehavi2011}.
The HOD model assigns each dark matter halo a probability of hosting $N$ galaxies, $P(N | M_{\mr{h}})$, that is dependent on the halo mass, $M_{\mr{h}}$. The galaxy population is divided into centrals, which are generally luminous galaxies that are located at or near the center of the halo, and satellites, less luminous galaxies which populate the outskirts of the halo. Each halo is allowed only one central but can have multiple satellites.
We follow the HOD parameterization of \cite{zehavi2011}: assuming a log-normal mass-luminosity distribution for central galaxies and a power-law distribution for satellite galaxies.
The expectation value for the number of galaxies for a luminosity thresholded sample (with absolute $r$-band magnitude $M_r<M_r^t$) is parameterized as
 \be
\begin{split}
\ensav{N(M_{\mr h}|M_r^t)} &= \left< N_c (M_{\mr h}|M_r^t)\right>\left(1+\left<N_s (M_{\mr h}|M_r^t)\right>\right) \\
                           &=\frac{1}{2}\left[1+\mr{erf}\left(\frac{\log M_{\mr h}-\log M_{\mr{min}}}{\sigma_{\mr{log}M}}\right)\right] \\
                           &\,\,\,\,\,\,\times\left[f_{\rm cen}+\left(\frac{M_{\mr h}}{M_1}\right)^{\alpha}\right]\,,
\label{eq:NM2}
\end{split}
\ee 
with model parameters $M_{\mr{min}}, M_1, \sigma_{\mr{log}M}, \alpha, f_{\rm cen}$.
For a DES simulation-based study using a similar HOD model, see \citet{park16}.
For simplicity we set the satellite cut-off scale of \citet{zheng2005} to zero, as it is not constrained by our data.

The central galaxy occupation is described by a softened step function with two parameters: (i) a transition mass scale $M_{\mr{min}}$, which is the halo mass at which the median central galaxy luminosity corresponds to the luminosity threshold, and (ii) a softening parameter $\sigma_{\mr{log}M}$ related to the scatter between galaxy luminosity and halo mass. The normalization of the satellite occupation function is $M_1$ and $\alpha$ is the high-mass slope of the satellite occupation function.
Finally, we introduce an additional parameter, $f_{\rm cen}$, the fraction of occupied halos, which allows us to relax the assumption that every halo above a certain mass contains a central galaxy.
Note that we have restricted ourselves to a simplified model in which $f_{\rm cen}$ is mass independent to reduce the dimensionality of our parameter space.
This parametrization (with $f_{\rm cen} = 1$) is able to reproduce the clustering of CFHTLS and SDSS galaxies over a large range of redshifts and luminosity thresholds \citep{zheng2005, zehavi2011}.
{\redmagic} galaxies are not quite complete to a luminosity threshold: they were selected using a cut in color space and this prompted the inclusion of $f_{\rm cen} $ as an additional free parameter. However, to obtain strong constraints on $f_{\rm cen}$, we would need to use the observed number density of galaxies as an additional constraint.
The priors used for each parameter are summarized in Table~\ref{tab:parameters}.
We choose priors based on earlier work with red galaxies: $f_{\rm cen}$ from redMaPPer \citep{rykoff14,rykoff16}; $0.01 < \sigma_{\mr{log}M} < 0.5$ and $0.4 < \log{M_1} - \log{M_{\rm min}} < 1.6$ from \citet{brown08}; $\alpha$ from \citet{white11} and \citet{parejko13}.
In addition, we checked that widening the priors for all parameters does not effect our main results in Sec.~\ref{sec:mass}.

Since the above prior choices are based on galaxy samples different from the DES \redmagic{} sample, we briefly discuss their applicability here.
Our fundamental assumption in adopting priors is that the principal impact of the galaxy selection is to set the mass scale $M_{\rm min}$ required to host a galaxy of the chosen minimum luminosity.
Thus, $M_{\rm min}$ is very clearly sample dependent.
The remaining parameters, however, we expect to be roughly comparable for different samples.
For instance, the scatter $\sigma_{\mr{log}M}$ ultimately reflects the scatter in central galaxy luminosity at fixed mass, so we expect $\sigma_{\mr{log}M}$ to be comparable for most galaxy samples.
Likewise, the ratio $M_1/M_{\rm min}$ is likely to be set by the mass $M_1$ required for a halo to host a substructure of mass $M_{\rm min}$, leading to the expectation $M_1/M_{\rm min}$ being roughly comparable for different luminosity or stellar mass thresholded samples.  Finally, the prior on the slope of the HOD is rather generous, 0.6-1.4, roughly a 40\% window around the naive expectation $\alpha=1$ corresponding to a constant galaxy/mass ratio.  Again, the critical point is that our principal results in Sec.~\ref{sec:mass} are insensitive to the details of these priors.

The HOD model allows us to predict, on average, the number of galaxies contained within each halo of a given mass. 
Together with the halo mass function, which tells us how many halos of each mass bin to expect within a given volume, and the halo model~\citep{Cooray02}, we can predict the expected clustering of these galaxies and their dark matter halos. 
On large scales, correlations between dark matter and galaxies in different halos dominates and can be approximated by the 2-halo matter-galaxy cross power spectrum as follows:
 \be
P_{\mr{gm}}^{2h}(k, z) \approx b_{\mr{gal}}P_{m}(k,z),
\ee 
where $P_{m}(k,z)$ is the linear dark matter power spectrum and the mean galaxy bias, $b_{\mr{gal}}$, is supplied by the HOD model as: 
 \be
\bar{b}_\mr{gal} = \frac{1}{\bar n _{\rm gal}}\int_0^\infty d M_{\mr h}\,\frac{dn}{dM_{\mr h}} b_{\mr h}(M_{\mr h})\ensav{N(M_{\mr h}|M_r^t)} \,, 
\label{eq:bg}
\ee 
where $\frac{dn}{d M_{\mr h}}$ is the halo mass function, $b_{h}(M_{h})$ is the halo bias relation and $\bar{n}_M$ is the mean number density of galaxies in the sample. We use the~\citet{Tinker:2008ff,Tinker:2010my} fitting functions for the halo mass function and halo mass--bias relation and the number density of galaxies can be calculated from the HOD as follows: 
 \be
\bar{n}_{\rm gal} = \int_0^\infty d M_{\mr h}\,\frac{dn}{dM_{\mr{h}}}\ensav{N(M_{\mr h}|M_r^t)}.
\label{eq:ngal}
\ee 

On smaller scales, the main contribution to the clustering is correlations between dark matter and galaxies in the same halo; this is described by the 1-halo term matter-galaxy cross power spectrum as a sum of central and satellite terms:
 \be
\begin{split}
P_{\mr{gm}}^{1h}(k,M_r^t) &= \frac{1}{\bar{\rho}_{\mr{m}}\bar{n}_{M}}\int d M_{\mr h} M_h \tilde u_{\mr h}(k, M_{\mr h})\frac{dn}{dM_{\mr h}}[\ensav{N_{\mr c}(M_{\mr h}|M_r^t)} \\
                      & + \ensav{N_{\mr s}(M_{\mr h}|M_r^t)}\tilde u_{\mr s}(k, M_{\mr h})]\,,
\end{split}
\ee 
where $\tilde u_{\mr{h}}$ is the Fourier transform of the halo density profile of mass $M_{\mr h}$, and $\tilde u_{\mr{s}}$ the Fourier transform of the spatial distribution of satellite galaxies in the halo.
We assume that the dark matter within a halo is distributed according to an NFW profile \citep{nfw1997}, with a \citet{b13} concentration-mass relation, $c (M) = 9 \times (1.686 / \sigma(M))^{-0.29} \times D(z)^{1.15}$,
where $\sigma(M)$ is the square root of the variance in a filter with mass $M$ and $D(z)$ is the growth factor at redshift $z$.
Note that we do not include subhalo 1-halo contributions from satellite galaxies in the model.
Given our conservative small-scale cutoff (see Appendix \ref{sec:deblending}) of 30 arcseconds and our statistical errors (see Fig.~\ref{fig:deltas-meas}) such a contribution is unnecessary.

Following~\cite{hayashi08} and \citet{zu14}, we transition between these two regimes by only taking the larger contribution of these two terms, such that
 \be \label{eq:pgm}
\xi_{\mr{gm}}(r,z) = 
\begin{cases}
\xi_{\mr{gm}}^{1h}(r,z) & \text{for } \;\;\; \xi_{\mr{gm}}^{1h}(r,z) \ge \xi_{\mr{gm}}^{2h}(r,z)\\
\xi_{\mr{gm}}^{2h}(r,z) & \text{otherwise}\\
\end{cases} 
\ee 
where $\xi_{\mr{gm}}^{1h}(r,z)$ is the 1-halo galaxy-matter cross correlation function where the galaxy and dark matter are both in the same halo, and $\xi_{\mr{gm}}^{2h}(r,z)$ is the 2-halo galaxy-matter cross correlation function describing correlations between a galaxy and dark matter in a different halo.

\begin{table}
	\centering
	\begin{tabular}{@{}lll@{}}
		\toprule 
		 & Prior range & Parameter \\
		\midrule
		$M_{\rm min}$ & 10.9 -- 13.6 & central halo mass \\
		$\sigma_{\mr{log}M}$ & 0.01 -- 0.5 & central galaxy HOD width \\
		$M_1$ & 13.3 -- 14.1 & mass of satellite's host \\
		$\alpha$ & 0.6 -- 1.4 & slope of the satellite distribution \\
		$f_{\rm cen}$ & 0 -- 0.45 & fraction of halos hosting a central galaxy \\
		$\sigma_8$ & 0.67 -- 0.93 & amplitude of clustering ($8 \mpch$ top hat) \\
		\bottomrule
	\end{tabular}
	\caption{HOD parameters and priors used in this work.}
	\label{tab:parameters}
\end{table}

Finally, we can relate the galaxy-mass power spectrum to our lensing observables.
For a single lens at redshift $z$ and source plane at $\zs$, we can write
 \be 
C_{g\kappa}(l | z, \zs) = \chi^{-2}(z)W_\kappa (z, \zs)P_\mr{gm}(k=l/\chi,z),
\label{cgkappa}
\ee 
where $\chi(z)$ is the comoving distance to redshift $z$, the lensing window function $W_\kappa (z, \zs)$ is
 \be \label{eq:window}
W_\kappa(z, \zs) = \frac{\bar{\rho}_m (z)}{(1+z)\Sigma_\mr{crit}(z, \zs)},
\ee 
and the critical surface density $\Sigma_\mr{crit}(z, \zs)$ of a flat universe is given by
 \be 
\Sigma_\mr{crit}^{-1}(z, \zs) = \frac{4\pi G}{1+z} \,\chi(z) \left[1-\frac{\chi(z)}{\chi(\zs)}\right] \, .
\ee 
We then transform $C_{g\kappa}(l)$ to real space and obtain the tangential shear $\gamma_\mr{t}(\theta)$:
 \be 
\label{gt_theory}
\gamma_t(\theta) = \int \frac{ldl}{2\pi}C_{g\kappa}(l)J_2(l\theta) \, .
\ee 
The tangential shear is related to the excess surface density, $\Delta\Sigma$, as
 \bea
\Delta \Sigma (R | z) & = & \bar{\Sigma}(<R) - \Sigma (R) \label{eq:ds_phys} \\
 & = & \Sigma_{\rm crit}(z, \zs) \, \gamma_{t}(R | z, \zs) \, , \label{ds_theory}
\eea 
where $\Sigma (R)$ is the surface mass density at the transverse separation $R$ from the center of the halo and $\bar{\Sigma}(<R)$ its mean within $R$.
Finally, we integrate over the redshift distribution of lens galaxies, $n(z)$, to obtain
 \be \label{ds_mean}
\left<\Delta \Sigma (R)\right> =  \frac{1}{\bar{n}} \int \de z \, n(z) \, \Delta \Sigma (R | z) \, .
\ee 
$\Delta\Sigma$ is a physical property of the lens and so does not depend on the source galaxies: this is clear from Eq.~(\ref{eq:ds_phys}) in which $\Delta\Sigma$ can be determined completely from the projected mass density $\Sigma$.
This independence from source redshifts is less obvious in Eq.~(\ref{ds_theory}), but note the $\Sigma_{\rm crit} (z, \zs)$ factor cancels with the same factor in Eq.~(\ref{eq:window}).

Note that throughout this paper we define the halo mass $M_{\rm h}$ as the mass within a sphere enclosing a mean density that is 200 times the mean mass density of the universe. (This mass is often labelled $M_{200m}$ in the literature.)
For modeling we fix cosmological parameters to $\Omega_m = 0.31$, $h = 0.67$, $\Omega_b = 0.048$, $n_s = 0.96$, and $w = -1$, all of which are consistent with the results of \citet{kwan17}.
We use physical length units throughout the paper.

%%%%%%%%%%%%%%%%%%%%%
\section{Data}
\label{sec:data}
%%%%%%%%%%%%%%%%%%%%%

The Dark Energy Survey is an ongoing wide-field multi-band imaging survey that will cover nearly 5000 square degrees of the Southern sky over five years.
The Dark Energy Camera
\citep[DECam,][]{flaugher15}
holds sixty-two 2048x4096 science CCDs, four 2048x2048 guider CCDs, and eight
2048x2048 focus and alignment chips, for a total of 570 megapixels covering a
roughly hexagonal footprint.  
Five filters are used during normal survey operations, \grizY.
The main survey will cover about 5000 square degrees in the South Galactic Cap
region, with approximately \nomepochs\ visits per field in the $r$, $i$ and
\zband s, for a $10\sigma$ limiting magnitude of about
$\sim$\nomdepth\ in the \iband.  

In this paper we use the largest contiguous region of Science Verification (SV) data which covers 139 square degrees with similar depth and filter coverage as the main DES survey. 
The SV data were taken during the period of November 2012 -- February 2013 
before the official start of the science survey.
All data used in this study is based on the DES SVA1 Gold catalog \footnote{\texttt{http://des.ncsa.illinois.edu/releases/sva1}} and several extensions to it.
The main catalog is a product of the DES Data Management (DESDM) pipeline version ``SVA1''.
The DESDM pipeline \citep{2006SPIE.6270E..23N, 2011arXiv1109.6741S, 
2012SPIE.8451E..0DM, desai2012, gruendl} begins with initial image processing on single-exposure images and astrometry measurements from 
the software package \textsc{SCAMP} \citep{2006ASPC..351..112B}. The single-exposure images were then stacked to produce 
co-add images using the software package \textsc{SWARP} \citep{2002ASPC..281..228B}. Basic object detection, 
point-spread-function (PSF) modeling, star-galaxy classification and photometry were done on the individual images as well 
as the co-add images using software packages \textsc{SExtractor} \citep{1996A&AS..117..393B} and \textsc{PSFEx} 
\citep{2011ASPC..442..435B}. 

For weak lensing we use the coadd images only for object detection, deblending,
fluxes (for use in \photoz\ measurements), and for the detailed informational
flags which are important for determining a good set of galaxies to use for shear measurement.
For the purposes of estimating galaxy shears, we instead perform object measurement on all
available {\SE} images in which an object was observed, using {\ME} fitting techniques \citep{jsz15}.

%%%%%%%%%%%%%%%
\subsection{Lens sample: redMaGiC}
\label{sec:magic}
%%%%%%%%%%%%%%%

The DES SV red-sequence Matched-filter Galaxy Catalog \citep[\redmagic,][]{rra15} is a catalog of photometrically
selected luminous red galaxies (LRGs).
Specifically, \redmagic\ 
uses the redMaPPer-calibrated model for the color of red-sequence galaxies as a function of magnitude and redshift \citep{rykoff14, rykoff16}.
This model is used to find the best fit photometric redshift for all galaxies irrespective of type, and the $\chi^2$ goodness-of-fit of 
the model is computed. For each redshift slice, all galaxies fainter than some minimum luminosity threshold $L_{\rm min}$ are 
rejected. In addition, \redmagic\ applies a $\chi^2$ cut $\chi^2 \leq \chi_{\rm max}^2$, where the cut $\chi_{\rm max}^2$ as a 
function of redshift is chosen to ensure that the resulting galaxy sample has a constant space density $\bar{n}$.
In this work, we use the sample with $\bar{n}=10^{-3} h^3 \rm{Mpc}^{-3}$; note that the redMaPPer algorithm assumes a flat cosmology with $\Omega_m=0.3$ in order to calculate the comoving density and luminosity distances.
We expect the result from our analysis to be only marginally sensitive to the cosmological parameters assumed.
The luminosity cut is $L\geq 0.5 L_*(z)$, where the value of $L_*(z)$ at z=0.1 is set to 
match the redMaPPer definition for SDSS.
The redshift evolution for $L_*(z)$ is that predicted using a simple passive evolution 
starburst model at $z=3$.
We utilize the \redmagic\ sample because of the excellent photometric redshifts of the \redmagic\ galaxy catalog: the \redmagic\ photometric redshifts are nearly unbiased, with a median bias $z_{\rm spec} - z_{\rm photo}$ of 0.005, scatter 
$\sigma_z/(1+z)$ of 0.017, and a $5\sigma$ redshift outlier rate of $1.4\%$. These photometric redshifts are used to split the \redmagic{} galaxies into different lens redshift bins in this study.
We refer the reader to \citet{rra15} for further details of this catalog. 

\begin{figure}
\centering
\resizebox{85mm}{!}{\includegraphics{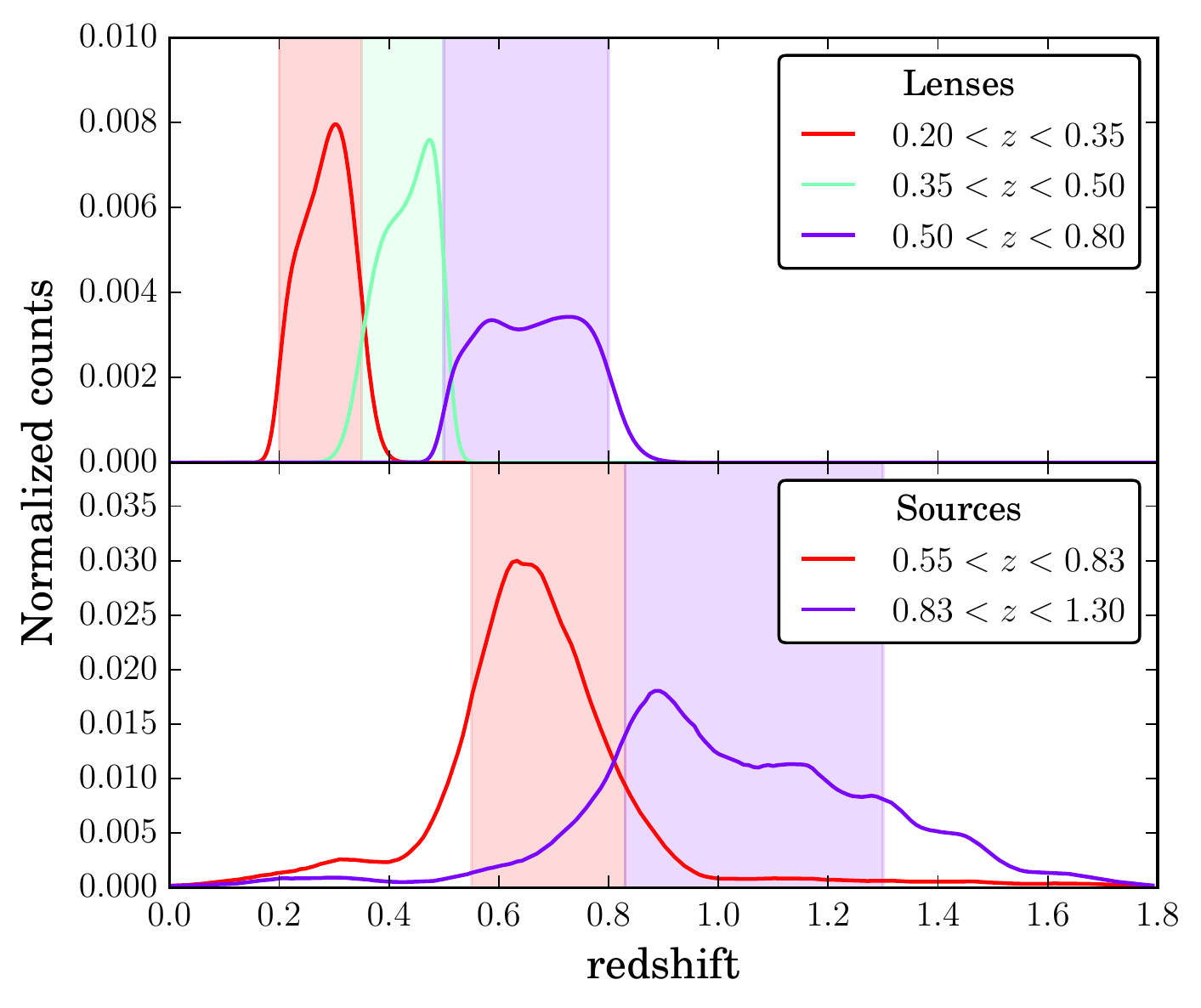}}
\caption{({\it top panel}): Redshift distributions of redMaGiC lens galaxies used in this work.
For lenses we show the stacked $N(z)$ from individual Gaussian distributions for each source. See text for details.
({\it bottom panel}): The same, but for our weak lensing source samples.
For sources we show the stacked $N(z)$ from individual SkyNet $p(z)$ distributions.
}
\label{fig:nzs}
\end{figure}

We use three redshift bins of lens galaxies, $0.2 < z < 0.35$, $0.35 < z < 0.5$, and $0.5 < z < 0.8$.
To plot the redshift distribution $N(z)$ of each bin in Fig.~\ref{fig:nzs}, we sum individual Gaussian redshift distributions for each lens, centered at the redMaGiC photo-$z$ estimate and with a width given by the redmagic photo-$z$ error.
Note that using a Gaussian for each lens neglects any outlier component: based on \citet{rra15} (especially Figures 10 and 11) the fraction of such outliers, and thus their effect on our modeling of the lens redshift distribution, is small.
Since the lens photometric redshifts have much higher precision than source redshifts we do not include a lens photo-$z$ systematic uncertainty contribution. However, an estimate of the true redshift distribution of lenses (as in Fig.~\ref{fig:nzs}) is taken into account in our modeling.
The number of lenses per redshift bin is $\sim 9000, 19000$, and $67000$ respectively.

%%%%%%%%%%%%%%%%%%%
%%%%%%%%%%%%%%%%%%%

%%%%%%%%%%%%%%%
\subsection{Source sample}
\label{sec:source}
%%%%%%%%%%%%%%%

Based on the SVA1 data, two shear catalogs were produced and tested extensively in \citet{jsz15} -- 
the \texttt{ngmix}~\footnote{{\texttt{https://github.com/esheldon/ngmix}}} \citep{s14}
and the \texttt{im3shape}~\footnote{\texttt{https://bitbucket.org/joezuntz/im3shape}} \citep{2013MNRAS.434.1604Z}
catalogs$^{\,1}$. The main results in our paper are based on \texttt{ngmix}, but we also cross-check 
with the \texttt{im3shape} catalog to demonstrate the robustness of our results. 

The \texttt{im3shape} implementation in this work estimates shapes by jointly 
fitting a parameterized galaxy model to all of the different single-exposure $r$-band images, finding the maximum 
likelihood solution.
The \textsc{PSFEx} software is used to fit pixelized models of the PSF, and those models are then interpolated to the galaxy positions.
Corrections to bias in the shear measurement associated with noise \citep{2012MNRAS.425.1951R, 2012MNRAS.427.2711K} 
are applied. The \texttt{im3shape} catalog has a raw number density of about 4.2 galaxies per square arcminute.

{\ngmix} uses sums of Gaussians to approximate common galaxy profiles: exponential disk, De Vaucouleurs' profile 
\citep{1948AnAp...11..247D}, and S{\'e}rsic profile \citep{1963BAAA....6...41S}.
Any number of Gaussians can be fit, either completely free or constrained to be co-centric and co-elliptical.
For the DES SV galaxy images, we used the exponential disk model. For the PSF fitting, an Expectation Maximization 
\citep{Dempster77maximumlikelihood} approach is used to model the PSF as a sum of three free Gaussians. 
Shear estimation was carried out using images in $r, i, z$ bands, which enabled a larger raw galaxy number 
density (6.9 galaxies per square arcminute). 
 
Photometric redshifts for source galaxies in DES-SV were studied in detail in \citet{bth15}, using 4 different photometric redshift codes (ANNz2, BPZ, SkyNet and TPZ). The details and capabilities of these codes on early DES data were 
already presented in \citet{2014MNRAS.445.1482S}, where they showed the best performance among a more extensive set of codes. For the results in this paper we use the photometric redshifts by SkyNet \citep{gfh14, bonnett15}, which is a neural network algorithm that was run on DES bands $griz$ and produced a probability density function (PDF) for the redshift of each galaxy in the DES-SV shape catalogs.
In addition to the source PDFs, we make use of a SkyNet point estimate of source photometric redshift given by the mean of the PDF.
Note that we only use SkyNet redshifts for the sources, as the redMaGiC algorithm produces a separate photo-$z$ estimate for each lens.

Our source redshift distributions are shown in Fig.~\ref{fig:nzs}.
Since our galaxy-galaxy lensing tests will provide validation for the cosmology results of \citet{kwan17}, that work uses these same redshift bins.
These bins are also consistent with those used in the DES SV results for cosmic shear \citep{btm15, des16}.
Note however that the $\Delta\Sigma$ estimator, according to the definition in Eq.~(\ref{eq:dsig_lens}), uses every background source behind each individual lens.

%%%%%%%%%%%
\subsection{Mock catalogs}
\label{sec:mock-catalog}

In Appendix~\ref{sec:sims} we make use of the ``Buzzard v1.0c'' version DES mocks to validate measurement covariances.
Three N-body simulations, a 1050 Mpc/h box with $1400^3$ particles, a 2600 Mpc/h box with $2048^3$ particles and a 4000 Mpc/h box with $2048^3$ particles, are combined along the line-of-sight to produce a full light cone out to DES depths.
These boxes were run with LGadget-2 \citep{springel05} and used 2LPTic initial conditions \citep{crocce06} with linear power spectra generated with CAMB \citep{lewis02}.
ROCKSTAR \citep{behroozi13} was used to find halos in the N-body volumes.
The ADDGALS algorithm \citep{busha13} is used to populate galaxies as a function of luminosity and color into dark matter only N-body simulations.
ADDGALS uses the relationship between local dark matter density and galaxy luminosity, determined by applying SHAM \citep{conroy06, reddick13} on a high-resolution tuning simulation, to populate galaxies directly onto particles in the low-resolution simulations.
Each galaxy is then assigned a color by using the color-density relationship measured in the SDSS \citep{aihara11} and evolved to match higher redshift observations.
For our mock lens sample, the same redMaGiC selection \citep{rra15} was run on the mock galaxies as on the data.
For the mock shears, galaxies are assigned sizes and shape noise using Suprime-Cam observations processed to match typical DES observing conditions \citep{sbd14}.
Finally, weak lensing shear for each source galaxy was computed using the multiple-plane lensing code CALCLENS \citep{becker13}.

%%%%%%%%%%%%%%%
\section{Measurement methodology}
\label{sec:method}
%%%%%%%%%%%%%%%

The most basic galaxy-galaxy lensing observable is the tangential shear of background source galaxies relative to the line joining the lens and source. For a given lens-source pair $j$ this is given by
 \be \label{eq:gammat}
\gamma_{t,j} = -\gamma_{1,j} \cos(2\phi_j) -\gamma_{2,j} \sin(2\phi_j) \, ,
\ee 
where $\gamma_{1,j}$ and $\gamma_{2,j}$ are the two components of shear measured with respect to a Cartesian coordinate system centered on the lens, and $\phi_j$ is the position angle of the source galaxy with respect to the horizontal axis of the Cartesian coordinate system.
However, the intrinsic ellipticity of individual source galaxies is much larger than the weak lensing shear, so it is necessary to average over many such lens-source pairs. For our measurements and null tests, we will compute the average in angular bins $\theta$ so that
 \be
\ave{\gamma_t^{\rm lens} (\theta)} = \frac{\sum_j w'_j \gamma_{t,j}}{\sum_j w'_j} \, ,
\ee 
where
 \be
w'_j = \frac{1}{\sigma_{\rm shape}^2 + \sigma_{{\rm m},j}^2} \, ,
\ee 
$\sigma_{\rm shape}$ is the intrinsic shape noise for each source galaxy, and $\sigma_{{\rm m},j}$ is the measurement error.
(This weight is the column labelled \texttt{w} in the shear catalogs described by \citealt{jsz15}.)
We use \verb+TreeCorr+\footnote{\texttt{https://github.com/rmjarvis/TreeCorr}} \citep{jbj04} to compute all galaxy-galaxy lensing measurements.

One advantage of this galaxy-shear cross-correlation over shear-shear correlations is that additive shear systematics (with constant $\gamma_1$ or $\gamma_2$) average to zero in the tangential coordinate system.
However, this cancellation takes place only when sources are distributed isotropically around the lens, an assumption that is not accurate near the survey edge or in heavily masked regions.
To remove additive systematics robustly we also measure the tangential shear around random points: such points have no net lensing signal, yet they sample the survey edge and masked regions in the same way as the lenses.
Our full estimator of tangential shear can then be written
 \be
\label{eq:gam}
\ave{\gamma_t (\theta)} = \ave{\gamma_t^{\rm lens} (\theta)} - \ave{\gamma_t^{\rm random} (\theta)} \, ,
\ee 
This measurement is directly comparable to the model prediction in eq. (\ref{gt_theory}). 

We will also find it useful to use another estimator that removes the dependence of the lensing signal on the source redshift. This will be especially helpful in carrying out null tests that involve splitting the source galaxy sample into two or more samples, then checking consistency between the measured lensing signal of each (see Sections~\ref{sec:split-size}, \ref{sec:split-snr}, \ref{sec:split-z}).
This observable is estimated from the measured shapes of 
background galaxies as
 \begin{equation} \label{eq:dsig_lens}
 \Delta\Sigma^{\rm lens}_{k}(R;\zl)=\frac{\sum_j 
\left[
w_j \gamma_{t,j}(R) / \Sigma_{{\rm crit}, j}^{-1}(\zl, \zs)
\right]
}{\sum_j w_j}
\end{equation} 
where 
the summation $\sum_j$ runs over all the background galaxies
in the radial bin $R$, around all the lens galaxy positions, 
and the weight for the $j$-th galaxy is given by
 \be \label{eq:weight}
w_j = w'_j \, \Sigma_{{\rm crit}, j}^{-2}(\zl, \zs) \, .
\ee 
Note that instead of $\theta$, we have binned source galaxies according to the radial distance $R$ in 
the region around each lens galaxy.
The weighting factor $\Sigma_{\rm crit}(\zl,\zs)$ is computed as a function of lens
and source redshifts for the assumed cosmology as
 \be \label{eq:sigma_crit}
\Sigma_{\rm crit} (\zl, \zs) = \frac{c^2}{4\pi G} \frac{D_A(\zs)}{D_A(\zl) D_A(\zl,\zs)} \, ,
\ee 
where $\Sigma_{\rm
crit}^{-1}(\zl,\zs)=0$ for $\zs<\zl$ and $D_A$ is the angular diameter distance.
We assumed a flat cosmology with $\Omega_m = 0.3$ when measuring $\Delta\Sigma$, although note that the results are not very sensitive to this value: the difference is well under 1\% when using $\Omega_m = 0.31$ (as in Sec.~\ref{sec:mass}).
Just as with the raw tangential shear, our final estimator involves subtracting the contribution around random points,
 \be
\label{eq:dsig}
\Delta\Sigma_{k}(R) = \Delta\Sigma^{\rm lens}_{k}(R) - \Delta\Sigma^{\rm random}_{k}(R) \, .
\ee 
We use 10 times as many random points as lenses so that noise from the random point subtraction is negligible.
We assign each random point a redshift drawn from the distribution of lens redshifts.
The measurement in Eq. (\ref{eq:dsig}) is directly comparable to the model prediction in Eq. (\ref{ds_mean}).
Note that in Eqs.~(\ref{eq:dsig_lens}-\ref{eq:dsig}) we use a point estimate of photometric redshift for each source (mean of the SkyNet $p(z)$, see Sec.~\ref{sec:source}).
However, we checked that the method of integrating over the full distribution of source redshifts \citep{scm12, nms12, msb2013} gives consistent results for both our central values and jackknife error bars.

We use a minimum fit scale $\sim 0.5$ arcminutes based on deblending constraints (see Appendix~\ref{sec:deblending}).
The maximum scale we use is 70 arcminutes, comparable to the size of our jackknife regions.
Using numerical simulations we have verified that the resulting jackknife covariance matrix estimate is accurate up to this scale (see Appendix~\ref{sec:sims} for details), above which the jackknife overestimates the errors from independent simulations.
Therefore including larger scales in the fits would still be conservative but there is very marginal gain in S/N so we elect to stop at 70 arcminutes.

While our photometric redshifts remove or downweight source galaxies near the lens redshift, some fraction of the sources will still be physically correlated with the lenses (e.g., \citealt{sjf2004, manetal06}).
This effect, corrected by applying ``boost factors,'' is most problematic near the center of massive halos where correlations are strongest.
Given our relatively high minimum fit scale of 0.5 arcminutes and small halo mass (compared to clusters, see Sec.~\ref{sec:mass}) boost factor corrections are negligible ($<1\%$) for our sample.

To estimate statistical errors, we divide the survey area into $N = 152$ spatial jackknife regions based on HEALpix\footnote{\texttt{http://healpix.sf.net}} \citep{ghb05}, each slightly smaller than 1 deg$^2$. We perform the measurement multiple times with each region omitted in turn. The covariance of the measurement \citep{nbg2009} is given by
 \be
\label{eq:cov}
C^{\rm stat}_{ij} = \frac{(N-1)}{N} \times \sum\limits_{k=1}^N \left[(\Delta \Sigma_i)^{k} - \overline{\Delta\Sigma_i}\right]
\left[(\Delta \Sigma_j)^{k} - \overline{\Delta\Sigma_j}\right]
\ee 
where $N$ is the number of jackknife regions, the mean value is
 \be \label{eq:avg}
\overline{\Delta\Sigma_i} = 
\frac{1}{N}
\sum\limits_{k=1}^N (\Delta\Sigma_i)^k\, ,
\ee 
and 
$(\Delta\Sigma_i)^k$
denotes the measurement from the $k$-th realization and the $i$-th
spatial bin.
Validation of this jackknife method using covariances from independent simulations is presented in Appendix~\ref{sec:sims}.
Finally, we apply the correction factor of \citet{hartlap07}, $\frac{N - N_{\rm bins} - 2}{N-1}$, to our inverse covariances when performing fits.
This factor is intended to correct for noise in the covariance matrix, but with $N = 152$ and a number of bins $N_{\rm bins} = 13$, this factor causes a very small change to our best-fit model and error bars shown in Sec.~\ref{sec:results}.

%%%%%%%%%%%%%%%
\begin{figure}
\centering
\resizebox{85mm}{!}{\includegraphics{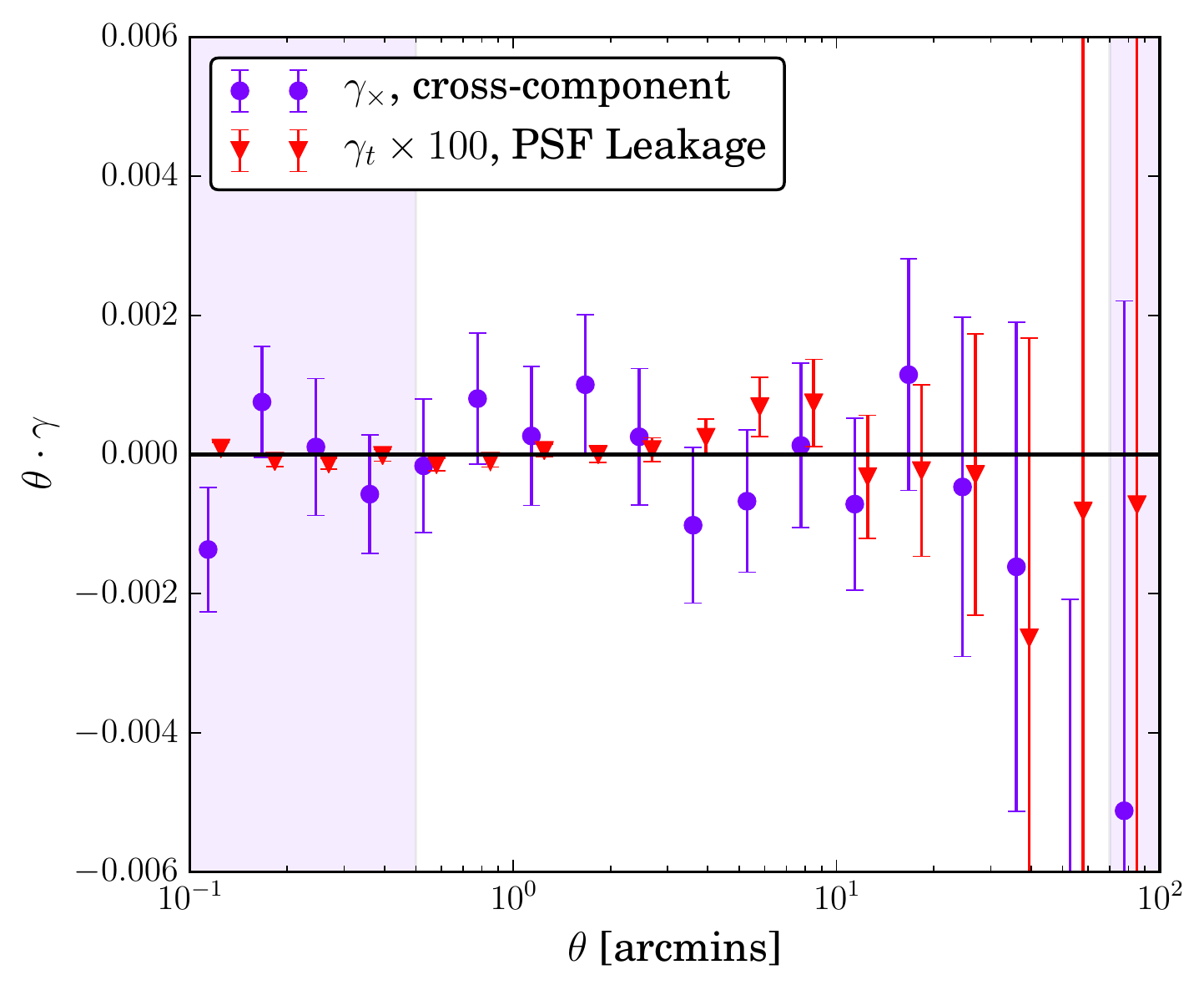}}
\caption{Cross-component of the shear (purple circles) and PSF leakage given by measuring tangential shear using the interpolated PSF at source positions (red triangles).
Scales $\theta < 0.5$ arcmin and $\theta > 70$ arcmin not used in our fits or $\chi^2$ results are indicated by shaded bands.
(Note that the pictured PSF leakage result is multiplied by 100, as labelled in the plot.)
Both tests pass with constant fits consistent with zero.
}
\label{fig:psf}
\end{figure}
%%%%%%%%%%%%%%%

%%%%%%%%%%%%%%%
\section{Data Tests}
\label{sec:tests}
%%%%%%%%%%%%%%%

We next perform a series of null tests used to check for and quantify the size of systematic uncertainties in our measurement due to biases in the shear and photometric redshift catalogs.
In order to quantify the comparison of the data tests in this section with the null hypothesis we compute the null $\chi^2$ for each of them in the following way:
 \begin{equation}
\chi_{\mathrm{null}}^2 = \sum_{i,j} \matr{y}^{T} (\matr{C}^{\rm stat})^{-1} \matr{y} \, ,
\end{equation} 
where $y_i$ corresponds to $\gamma_i$ or $\Delta\Sigma_i$, and $C^{\rm stat}_{ij}$ is the corresponding covariance matrix.
A list of the null $\chi^2$ value for each of the tests can be found in Table~\ref{tab:syst}.
While $\chi^2$ is a helpful statistic, it is possible that two tests, each with the same $\chi^2$ values, could indicate different levels of systematic error.
An extreme example is the case where every data point is positive for one test, whereas half the points are positive and half are negative (with the same amplitude) for the second test.
Thus for each test we also quote the result of a single parameter, constant fit to the data (including the full jackknife covariance).
Since we perform many tests, some fraction of them are expected to differ from zero simply by chance.
For example, even in the absence of systematics, 1 out of 3 tests will differ from zero by more than 1$\sigma$, and 1 out of 25 will differ by more than 2$\sigma$.
Our criteria for declaring a test as ``passed'' is that the constant fit should be within $2\sigma$ of zero, but note that using $1.5\sigma$ instead does not change the status of any tests.

In this section we present tests that are unique to galaxy-galaxy lensing.
Other tests of the shear catalog, mostly focused on validation of cosmic shear results, have been presented in \citet{jsz15} and \citet{btm15}.
For the tests that were already studied in prior DES SV work, we simply summarize the implications for our galaxy-galaxy lensing measurements in Appendix \ref{sec:syst-other}.
These tests include deblending and sky subtraction (\ref{sec:deblending}),
multiplicative shear bias (\ref{sec:shear-bias}), and stellar contamination and shear around stars (\ref{sec:stars}).
For most tests in this section we show only the result with \texttt{ngmix}, but the results for both shear pipelines are summarized in Table~\ref{tab:syst}.

%%%%%%%%%%%%%%%
\subsection{Cross-component, PSF leakage, and random point shear}
%%%%%%%%%%%%%%%

For the first three tests (lensing cross-component, PSF leakage, and shear around random points) we show only results using lenses between $0.35 < \zl < 0.5$ and sources $0.83 < \zs < 1.3$.
This range is emphasized because it is the fiducial lens bin of \citet{kwan17}.
Note however, these three tests also pass with the other lens and source bin combinations.

In the cross-component test we measure the cross shear around lens galaxies, which is a 45$^{\circ}$ rotated signal with respect to the tangential shear defined in Eq.~(\ref{eq:gammat}). This signal should be compatible with zero if the shear is only induced by gravitational lensing, and therefore provides a test for systematic errors related to PSF correction, which can leak into both tangential and cross components of the galaxy shear. In Fig.~\ref{fig:psf} we show the resulting cross-shear measured around redMaGiC lenses.
The reduced $\chi^2 = 8.3 / 13 \, (10.9/13)$ for \texttt{ngmix} (\texttt{im3shape}), and the test is consistent with a constant fit equal to zero.

%%%%%%%%%%%%%%%
\begin{figure}
\centering
\resizebox{85mm}{!}{\includegraphics{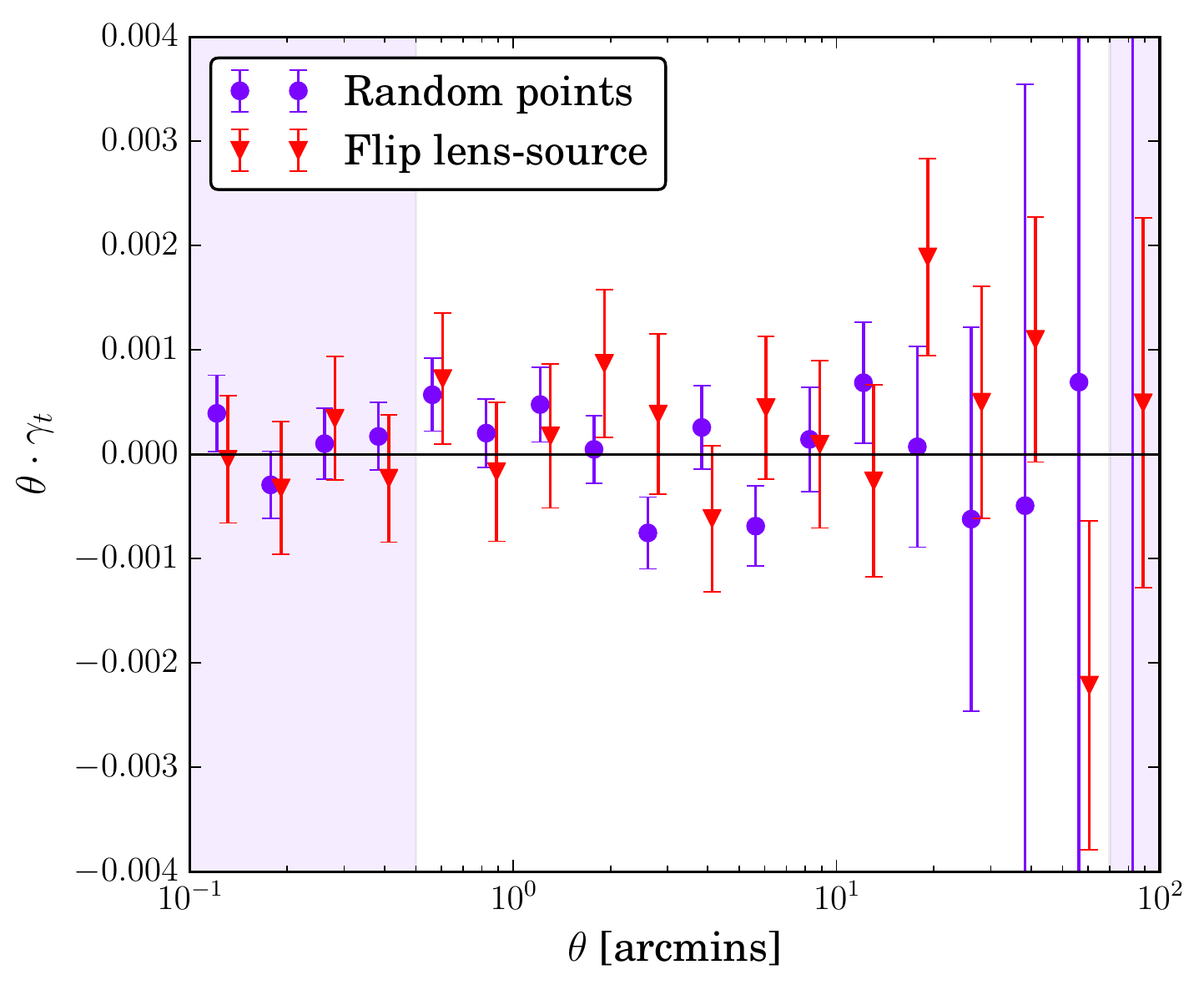}}
\caption{Same as Fig.~\ref{fig:psf}, but showing tangential shear around random points (purple circles) and tangential shear of foreground sources (red triangles).
Both tests pass easily (see detailed numbers in Table~\ref{tab:syst}).}
\label{fig:random-flip}
\end{figure}
%%%%%%%%%%%%%%%

Measuring source galaxy shapes requires modeling them convolved with the PSF pattern imprinted by the atmosphere and optics.
However, this process is imperfect, such that $\sim 1\%$ ($3\%$) of the PSF shape may ``leak'' into the measured galaxy shape for \texttt{ngmix} (\texttt{im3shape}), based on tests in \citet{jsz15}.
Note that while \citet{jsz15} found leakage is consistent with zero for both pipelines, the values we quote above conservatively assume the maximum allowed leakage within the $1\sigma$ errors of \citet{jsz15}.
In order to quantify this systematic, we measure the tangential shear of the PSF interpolated to the source galaxy locations, where again the tangential shear is measured around the redMaGiC lenses.
In Fig.~\ref{fig:psf} we show the result, multiplied by a factor of 100.
It is consistent with zero, and furthermore given the small upper bounds on the leakage, even these small fluctuations about zero are much smaller than our measured lensing signal (see Sec~\ref{sec:measure}).

While our estimator of galaxy-galaxy lensing in Eq.~(\ref{eq:dsig}) involves subtracting the signal around random points that trace the same survey geometry, it is nonetheless useful to confirm that this correction is small at all scales used in the analysis.
This measurement tests the importance of systematic shear which is especially problematic at the survey boundary, and allows us to compare the magnitude of the systematic shear with the magnitude of the signal around actual lens galaxies.
In Fig.~\ref{fig:random-flip} we show the result, which is consistent with the null hypothesis.
Again, see Table~\ref{tab:syst} for all the detailed test results.

%%%%%%%%%%%%%%%%%%%
\begin{figure*}
\centering
\resizebox{180mm}{!}{\includegraphics{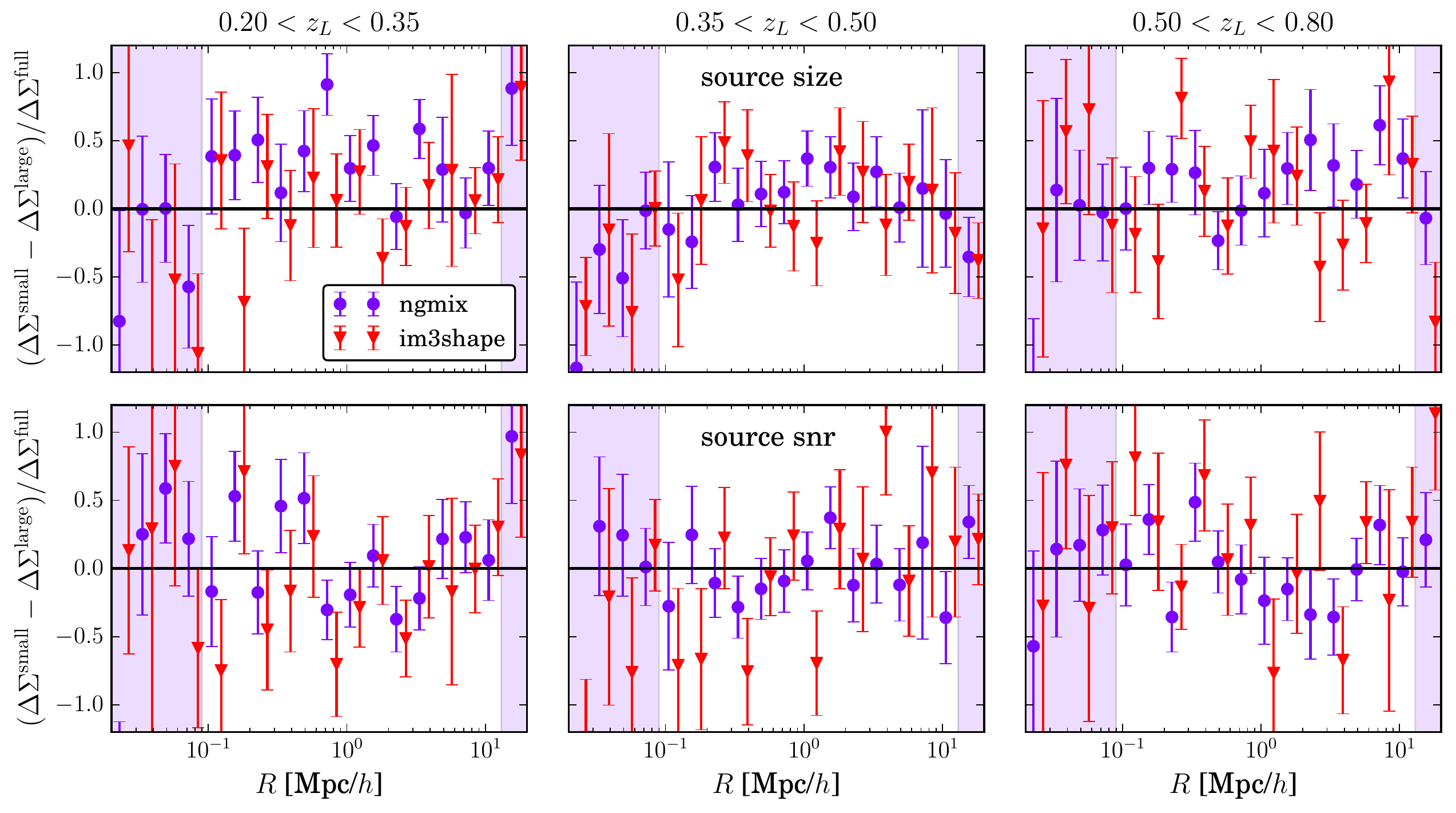}}
\caption{({\it upper panels}): The difference between the $\Delta\Sigma$ signals from small and large source galaxies relative to the central values of the full sample for $\texttt{ngmix}$ (purple circles) and $\texttt{im3shape}$ (red triangles), measured for all three lens samples.
({\it lower panel}): The same, but showing results for the split between low and high signal-to-noise sources.
These splits are not always consistent with zero, see the text in Sec.~\ref{sec:split-size} for a detailed discussion.
Only data in the middle, unshaded region of each plot is used for our tests and measurements.
}
\label{fig:size-snr}
\end{figure*}
%%%%%%%%%%%%%%%%%%%

%%%%%%%%%%%%%%%
\subsection{Flip lens and source samples}
%%%%%%%%%%%%%%%

In reality, only source galaxies which are {\it behind} foreground lenses will be lensed. However, in the presence of redshift errors some low redshift sources will mistakenly be put behind the lenses, and vice-versa. To get some handle on this effect, we repeat the measurement of $\ave{\gamma_t}$ but using foreground sources ($0.2 < \zs < 0.5$) stacked around our highest redshift lenses ($0.5 < \zl < 0.8$).
Note that this test is noisy and thus easily satisfied when applied to the lower and medium redshift lens samples due to insufficient numbers of foreground sources.
The result is shown in Fig.~\ref{fig:random-flip} and Table~\ref{tab:syst}, and is consistent with the null hypothesis.

%%%%%%%%%%%%%%%
\subsection{Source size splits}
\label{sec:split-size}
%%%%%%%%%%%%%%%

Shape measurements may be more biased for source galaxies which are smaller and less well resolved.
Although we have applied multiplicative bias corrections to our measurements (for \texttt{im3shape}) or checked they are small (for \texttt{ngmix}) (see \citealt{jsz15}), we test to ensure that there is no residual bias by splitting the source galaxies into two samples with different size.
We use the ``round'' measure of size \citep{jsz15} for \texttt{ngmix}, \texttt{exp\_T\_r}, splitting the two samples at 0.45 and measuring their difference relative to the central values of the combined sample.
For \texttt{im3shape}, the corresponding cut is at 1.4 using the $R_{\rm gpp}/R_{\rm p}$ size parameter, where $R_{gpp}/R_p$ is the full width at half maximum (FWHM) of the convolved model divided by the FWHM of the PSF for each exposure.
We use these values in order to make a 60\% / 40\% split of the source galaxies, with slightly more sources in the smaller size bin.
This choice is made since smaller sources may be somewhat more noisy, but the test results are not strongly dependent on the exact cut.

\begin{figure}
\centering
\resizebox{85mm}{!}{\includegraphics{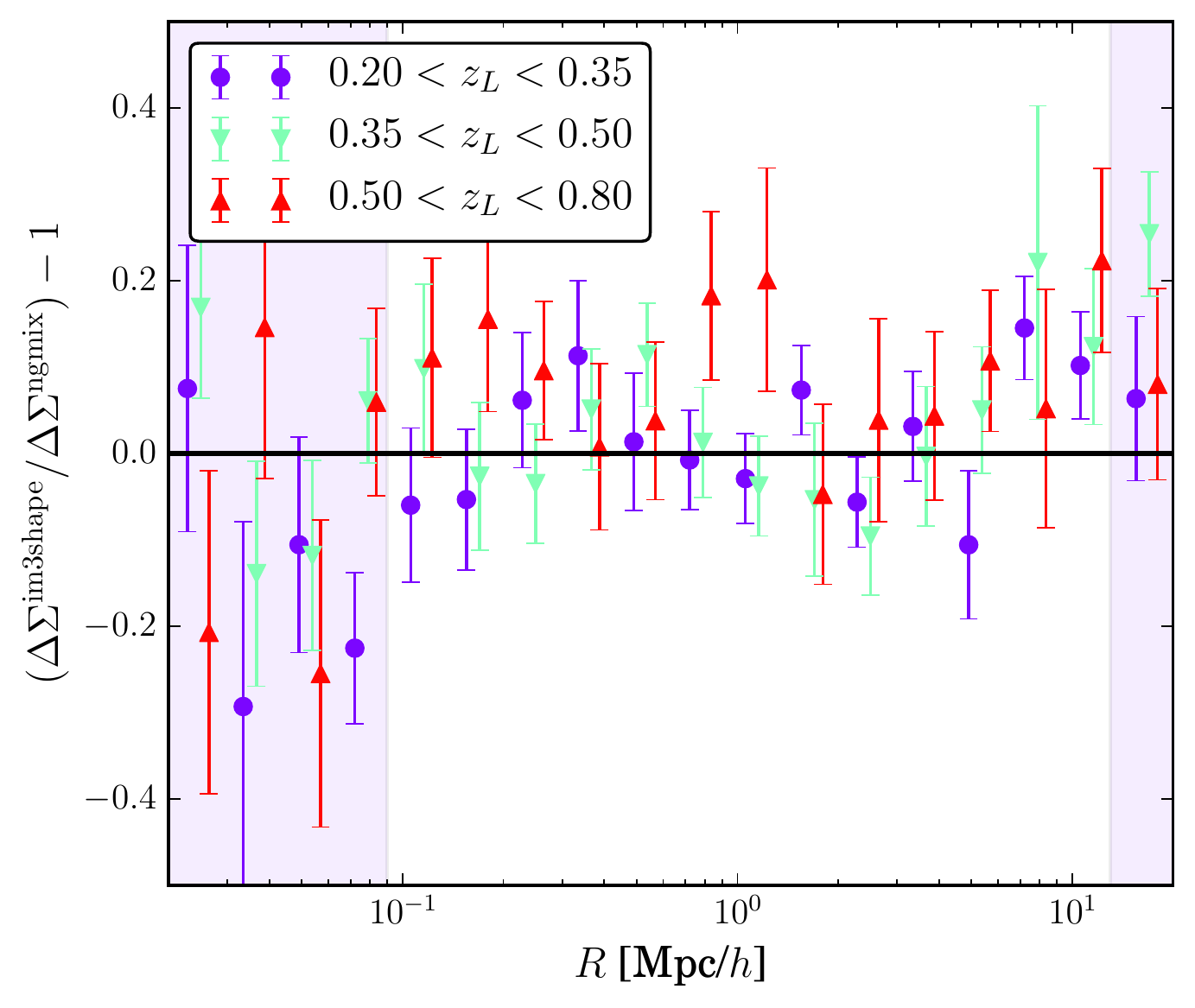}} \\
\caption{Ratio of \texttt{im3shape} and \texttt{ngmix} (our fiducial pipeline) $\Delta\Sigma$ measurements, for all lens redshift bins.
For the two lower redshift bins, the difference between shear pipelines is very small at about 1\%.
The higher redshift bin is more discrepant at 9\%, although this difference is still within our $1\sigma$ errors.
}
\label{fig:deltas-comp}
\end{figure}

The result for the lowest redshift lens bin, shown in Fig.~\ref{fig:size-snr}, has a reduced $\chi^2 = 39.4/13 \, (7.6/13)$ for \texttt{ngmix} (\texttt{im3shape}).
This is a very high $\chi^2$ result for \texttt{ngmix}.
As described at the beginning of this Sec.~\ref{sec:tests}, to quantify any possible systematic uncertainty we fit a constant to the fractional difference to determine its magnitude.
The result is a constant $= 0.36 \pm 0.08$, thus the difference in the shear of small and large source galaxies is significant at about $4-5\sigma$.
We assume that the true shear induced by lensing falls somewhere between the answer given by small and large sources.
The best-case scenario is that the true shear falls exactly between the two, in which case our full sample would have an unbiased average shear.
The worst case scenario is that either small or large sources give biased estimates of the true shear while the other is unbiased.
In this case the bias of the full sample is half the constant fit above, $0.36 / 2 = 18\%$.
However, this scenario is at odds with other tests, including the fact that \texttt{ngmix} and \texttt{im3shape} shears are in close agreement.
In Fig.~\ref{fig:deltas-comp} we show the measured $\Delta\Sigma$ ratio between the two shear pipelines.
The consistency is excellent for the two lower lens redshift bins, with systematic differences of $\sim 1\%$ for the two lower lens redshift bins.
Even for the highest lens bin, which relies on the highest redshift sources, the difference is only 9\%.
The good agreement between pipelines provides some evidence for the smaller estimate of systematic uncertainty.
In Table~\ref{tab:syst} we note the total systematic uncertainty for \texttt{ngmix}, both with and without this size split.
When performing HOD fits in Sec.~\ref{sec:mass} we do not include this 18\%.

The picture is significantly better for \texttt{ngmix} when using the middle and highest lens redshift bins.
These results are also shown in Fig.~\ref{fig:size-snr} and indicate a conservative $\sim 2\sigma$ systematic of 7\% (9\%) for the size split of middle (highest) lens redshift bin.
These two lens bins use nearly the same sources, but weighted differently according to Eq.~(\ref{eq:weight}).
We have not been able to identify the source of the size split difference, but note here another possibility for investigation.
A size-filtered subsample might have a redshift distribution that is different from what was estimated based on $g,r,i,$ and $z$ magnitudes alone.
For example, \citet{gruen14} found an $\sim$ 5--10\% effect in the the mean $D_A(\zl,\zs)/D_A(\zs)$ from a size split (between a large size subset and the full sample) selected in B,R, and I filters.
This means that given the same color and magnitude -- and therefore the same implied redshift -- the difference in the shear between the large sources and the full sample was $\sim$ 5--10\%.
Similarly \citet{applegate14} found a 5\% difference in the lensing signal when using a large size subset.
We do not have the resources to explore this effect further in DES-SV, but it will be worth studying in DES Year 1 data.

%%%%%%%%%%%%%%%%%%%
%%%%%%%%%%%%%%%%%%%

%%%%%%%%%%%%%%%
\subsection{Source S/N splits}
\label{sec:split-snr}
%%%%%%%%%%%%%%%

While source galaxy S/N is partially correlated with size, it is a distinct parameter that may separately influence the accuracy of fitted shapes.
Thus, we find the difference of two samples with S/N $< 45$ and S/N $> 45$, where S/N is the \texttt{ngmix} ``round'' signal-to-noise measure \texttt{exp\_s2n\_r}.
For \texttt{im3shape}, source S/N is given by \texttt{snr} in the \texttt{im3shape} catalog.
(See \citealt{jsz15} for more details on these measurements of galaxy S/N.)
Again this split puts 60\% of the sources into the smaller S/N bin.
The result is shown in Fig.~\ref{fig:size-snr} and summarized in Table~\ref{tab:syst} for both pipelines.
While the size split in Sec.~\ref{sec:split-size} failed for \texttt{ngmix} and passed for \texttt{im3shape}, here the trend is reversed.
The constant fit to the difference is consistent with zero for \texttt{ngmix}, but indicates a systematic uncertainty with magnitude $\sim 10\%$ for \texttt{im3shape}, using the lowest lens redshift bin.
The middle and high lens redshift bins pass this test for both pipelines as summarized in Table~\ref{tab:syst}.

Note that when repeating this test for the highest redshift lens bin, we adjust the cut between low and high S/N samples to 35.
This is necessary because for both catalogs the source S/N distribution is significantly different for small and large S/N galaxies.
The adjusted cut ensures the number of galaxies in the small S/N sample remains $\sim 60\%$.

%%%%%%%%%%%%%%%
\subsection{Source redshift splits}
\label{sec:split-z}
%%%%%%%%%%%%%%%

\begin{figure}
\centering
\resizebox{85mm}{!}{\includegraphics{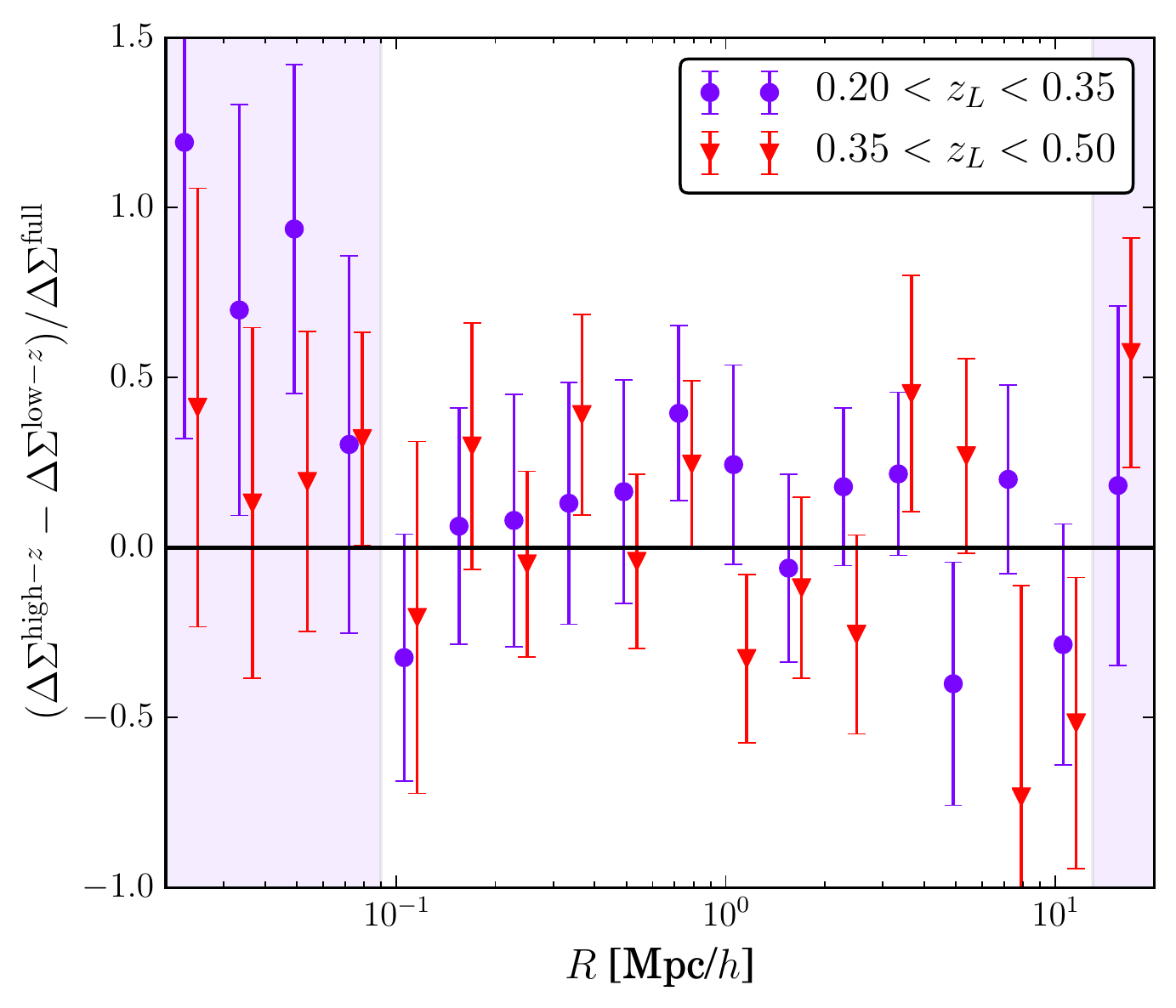}}
\caption{Same as Fig.~\ref{fig:size-snr} but showing the effect of splitting the \texttt{ngmix} sources into medium ($0.55 < \zs < 0.83$) and high redshift ($0.83 < \zs < 1.30$) bins.
The result is shown for two lens redshift bins, $0.2 < \zl < 0.35$ (purple circles) and $0.35 < \zl < 0.5$ (red triangles).
It is consistent with zero for both lens redshift bins.
}
\label{fig:zsplit}
\end{figure}

For the following null test, we look for differences in the lensing signal computed using two source samples split on redshift.
For continuity, these are the two higher redshift bins used by \citet{btm15}.
The bins are $0.55 < z < 0.83$, and $0.83 < z < 1.30$ where $z$ is the mean of the source SkyNet $p(z)$.
We compute the difference of $\Delta \Sigma (R)$ for both samples, for both the low and medium redshift lens bins.
The result is shown in Fig.~\ref{fig:zsplit}, and for both \texttt{ngmix} and \texttt{im3shape} shears the result is consistent with zero for both lens samples.
For \text{ngmix} the lowest redshift lens bin has a constant fit $5\pm4$\%, which is outside $1\sigma$.
However as described at the beginning of Sec.~\ref{sec:tests}, one in three independent tests are expected to fail at this level.
As with previous sections, the $\chi^2$ and constant fit numbers are described in detail in Table~\ref{tab:syst}.

%%%%%%%%%%%%%%%%%%%
%%%%%%%%%%%%%%%%%%%

%%%%%%%%%%%%%%%%%%%%%
\begin{table*}
Table~\ref{tab:syst}: $\chi^2$ for data and other tests and resulting systematic uncertainties \\
\begin{tabular}{l c | r r c | r r c}
\hline
 & & \multicolumn{3}{c}{\texttt{ngmix}} & \multicolumn{3}{c}{\texttt{im3shape}} \\
\hline
Test & lens redshift & $\chi^2 /$ ndf & Systematic uncertainty & passed & $\chi^2 /$ ndf & Systematic uncertainty & passed \\
\hline
cross-component & medium & 8.3 / 13 & $(6.7 \pm 5.1) \times 10^{-5}$ & \ding{51} & 10.9 / 13 & $(6.0 \pm 7.8) \times 10^{-5}$ & \ding{51} \\
PSF leakage & medium & 11.0 / 13 & $(0.1 \pm 3.5) \times 10^{-5}$ & \ding{51} & 13.3 / 13 & $(1.5 \pm 1.1) \times 10^{-5}$ & \ding{51} \\
random points & medium & 16.4 / 13 & $(0.4 \pm 2.8) \times 10^{-5}$ & \ding{51} & 7.2 / 13 & $(2.0 \pm 3.6) \times 10^{-5}$ & \ding{51} \\
flip lens-source samples & high & 12.0 / 13 & $(0.6 \pm 1.8) \times 10^{-5}$ & \ding{51} & 8.7 / 13 & $(1.2 \pm 1.9) \times 10^{-5}$ & \ding{51} \\
\hline
source size splits & low & 39.4 / 13 & up to $18\pm4$\% & $\times$ & 7.6 / 13 & up to $3\pm5$\% & \ding{51} \\
source S/N splits & low & 13.2 / 13 & up to $1\pm4$\% & \ding{51} & 18.0 / 13 & up to $10\pm5$\% & $\times$ \\
source size splits & medium & 9.1 / 13 & up to $7 \pm 3$\% & $\times$ & 8.7 / 13 & up to $5\pm5$\% & \ding{51} \\
source S/N splits & medium & 8.8 / 13 & up to $4\pm4$\% & \ding{51} & 16.1 / 13 & up to $3\pm6$\% & \ding{51} \\
source size splits & high & 13.9 / 13 & up to $9\pm4$\% & $\times$ & 23.6 / 13 & up to $5\pm5$\% & \ding{51} \\
source S/N splits & high & 13.2 / 13 & up to $0\pm4$\% & \ding{51} & 19.7 / 13 & up to $6\pm6$\% & \ding{51} \\
source redshift splits & low & 8.8 / 13 & up to $5\pm4$\% & \ding{51} & 14.6 / 13 & up to $4\pm4$\% & \ding{51} \\
source redshift splits & medium & 11.7 / 13 & up to $0\pm4$\% & \ding{51} & 12.3 / 13 & up to $1\pm5$\% & \ding{51} \\
\hline
intrinsic alignments & low, medium & & 2\% & & & 2\% & \\
intrinsic alignments & high & & 3\% & & & 3\% & \\
residual multiplicative bias & all & & 2\% & & & 1\% & \\
stellar contamination & all & & 2\% & & & 2\% & \\
tangential shear around stars & all & & consistent with 0 & & & consistent with 0 & \\
\hline
\hline
\textbf{Total} & low & & 4\% (19\%) & & & 3\% (10\%) & \\
 & medium & & 4\% (8\%) & & & 3\% (3\%) & \\
 & high & & 4\% (10\%) & & & 4\% (3\%) & \\
\end{tabular}
\caption{Summary of all test results described in \S~\ref{sec:tests}.
We show reduced $\chi^2$ and constant fit results for all data tests presented in this paper.
For the tests based on splitting the sample into two halves, the number in the table is an upper bound on the systematic uncertainty in $\Delta\Sigma$ (see the discussion in \S~\ref{sec:split-size}).
For brevity, the first three lines only show results for one lens bin, but the other bins are also consistent with zero.
The last three rows show the net systematic uncertainty for each lens bin, obtained by adding in quadrature the systematic uncertainties from individual tests.
In each case, the total error is shown both without and with (in parentheses) the inclusion of the results of the source size and S/N split.
}
\label{tab:syst}
\end{table*}

%%%%%%%%%%%%%%%%%%%%%%
\subsection{Intrinsic alignments}
%%%%%%%%%%%%%%%%%%%%%%

We have so far assumed that a source galaxy's observed tangential ellipticity is an unbiased estimate of its tangential shear. This is valid if the source galaxy's intrinsic ellipticity is not correlated with the direction to the lens, which is reasonable if source and lens galaxies are separated in redshift, i.e., not physically close. However since we have only imperfect, photometric redshift estimates, there is some overlap in redshift between sources and lenses (see Fig.~\ref{fig:nzs}).

The intrinsic shapes of galaxies are correlated with the cosmological density field, an effect known as ``intrinsic alignments'' (IA). Thus, for lens-source pairs which are physically close, source galaxies may be preferentially aligned with the direction to the lens galaxy.
For example, in the commonly used linear alignment model \citep{ca01, hs04}, the intrinsic ellipticity is linearly related to the tidal field (with a free amplitude), producing a correlation between intrinsic ellipticity and density that has the same scale dependence as the shear on linear scales.
On large (2-halo) scales, the linear alignment model is expected to describe elliptical source galaxies well, especially when nonlinear contributions are included, such as in the ``nonlinear linear alignment model'' \citep{bk07} or the ``complete tidal alignment model'' \citep{blazek15}.
This has been confirmed by measurements of LRG alignment (e.g., \citet{sm14}).

Accounting for the full photometric redshift distributions, we find that for our $0.35 < z < 0.5$ lens bin and $0.55 < z < 0.83$ source bin, the nonlinear linear alignment model predicts at most an $\sim 4\%$ contamination of the tangential shear signal.
The 4\% results from using the fiducial IA amplitude ($A=1$) from \citet{bk07}.
(See Figure 8 of \citet{des16}, which estimated IA amplitude for this source sample for different model scenarios. The model with the largest value of $A$ found $A = 2 \pm 1$, but the result was highly model dependent with some scenarios consistent with 0.)
Roughly half of the S/N of the $0.35 < z < 0.5$ lens bin comes from sources in the range $0.55 < z < 0.83$; based on the $N(z)$ in Fig.~\ref{fig:nzs} the higher redshift sources will not overlap in redshift with the lenses.
Thus we estimate a 2\% intrinsic alignment contamination of our measurement for this lens bin.
This contamination is likely to be reduced further when using the $\Delta \Sigma (R)$ statistic, which downweights sources with redshift close to the lens.
The lower lens bin $0.2 < z < 0.35$ will have less contamination (see Fig.~\ref{fig:nzs}), so we conservatively use 2\% for this bin as well.
Repeating the above calculation for the $0.5 < z < 0.8$ lens bin and $0.83 < z < 1.3$ source bin, we find a 3\% contamination.
For each lens bin we add the estimated IA contamination in quadrature to our other sources of error.

As an additional check, we compare the 2\% estimate for the $0.35 < z < 0.5$ lenses and $0.55 < z < 0.83$ sources to the most relevant current observational constraints of \citet{bm12}. Using an SDSS DR7 LRG lens sample and photometric source sample, \citet{bm12} solve simultaneously for the intrinsic alignment and lensing signals. They find model-independent upper limits (95\% confidence level) on the contamination of $\Delta \Sigma (R)$ of $\sim6\%$  for a projected separation $1 h^{-1}\textrm{Mpc}$. This is further reduced to $\sim3\%$ when assuming that blue source galaxies have zero intrinsic alignment amplitude. Thus, the 2\% estimated using our specific lens and source redshift distributions is compatible with previous observational constraints. Although it is beyond the scope of this work, the approach in \citet{bm12} of constraining the intrinsic alignment signal simultaneously with the lensing signal should be pursued in future DES analyses with improved statistical power.

%%%%%%%%%%%%%%%%%%%%%%
\subsection{Non-weak shear and magnification}
%%%%%%%%%%%%%%%%%%%%%%

The observable reduced shear will differ from $\gamma_{t}$ according to
 \be
g_t = \frac{\gamma_t}{1 - \kappa} \, .
\ee 
Since $\kappa$ is always rising with decreasing distance from the halo center, the error from using $\gamma_t$ rather than $g_t$ will be highest at our lowest fit radii, $R\sim 0.1 \mpch$.
Taking our largest best-fit halo mass $\sim 2\times 10^{13} M_\odot / h$ from Fig.~\ref{fig:hod_z_evol}, and assuming an NFW profile for $\kappa$, we find that the fractional difference between the shear and reduced shear $(\gamma_t - g_t)/g_t$ is at most 3.5\%.
The difference falls to 2\% by our second data point at $R \lesssim 0.13$.
At a halo mass $3\times 10^{13} M_\odot / h$, roughly the upper edge of the most massive 1-sigma constraints in Fig.~\ref{fig:hod_z_evol}, the difference at the lowest fit radius is at most 5\%.
Since the error in ignoring non-weak shear effects is much less than our other sources of systematic and statistical error, we neglect it in the analysis.
Similar to non-weak shear, magnification is a potential systematic effect that is more important for lenses with larger $\kappa$ than our sample.
See \citet{manetal06} for a galaxy-galaxy lensing specific discussion of the effects of magnification.

%%%%%%%%%%%%%%%%%%%%%%
\subsection{Total systematic uncertainty budget}
%%%%%%%%%%%%%%%%%%%%%%

All sources of systematic uncertainty studied in this paper are summarized in Table~\ref{tab:syst}.
This list should account for all the important systematic uncertainties in our measurements.
The final lines of Table~\ref{tab:syst} show the net systematic uncertainty for each lens redshift bin, obtained by adding the individual sources of systematic uncertainty in quadrature.
The systematic difference between large and small sources, photo-$z$ bias, shear calibration, and stellar contamination all cause multiplicative biases on $\Delta\Sigma$.
Thus we estimate the systematic covariance matrix for each lens bin as $C^{\rm syst}_{ij} = f^2 \times \Delta\Sigma_i \Delta\Sigma_j$, where $f$ is the total systematic uncertainty for that lens bin in Table~\ref{tab:syst} (e.g., 4\% for the \texttt{ngmix} shears and middle-z redMaGiC sample), and $\Delta\Sigma(R_i)$ is abbreviated $\Delta\Sigma_i$.
In Table~\ref{tab:syst} we show results both with and without the size and S/N splits, as discussed in Sec.~\ref{sec:split-size}.
Our total covariance matrix used in the HOD fits (Sec.~\ref{sec:mass}) is then $C_{ij} = C^{\rm stat}_{ij} + C^{\rm syst}_{ij}$, where we drop the size and S/N splits.
This approach is similar to that followed for galaxy-galaxy lensing measurements by, e.g., \citet{manetal06}.
That work folded together shear calibration, photo-$z$ bias, and stellar contamination into a systematic uncertainty which was added in quadrature to the statistical errors when performing fits to the halo mass.

%%%%%%%%%%%%%%%%%%%%%%%%%%%%
\section{Results}
\label{sec:results}
%%%%%%%%%%%%%%%%%%%%%%%%%%%%

Having carried out a number of successful null tests and quantified the remaining systematic uncertainties, in this section we present the galaxy-galaxy lensing signal and the best-fit mean halo mass.

%%%%%%%%%%%%%%%
\subsection{Measurement}
\label{sec:measure}
%%%%%%%%%%%%%%%

%%%%%%%%%%%%%%%%%%%
\begin{figure}
\centering
\resizebox{85mm}{!}{\includegraphics{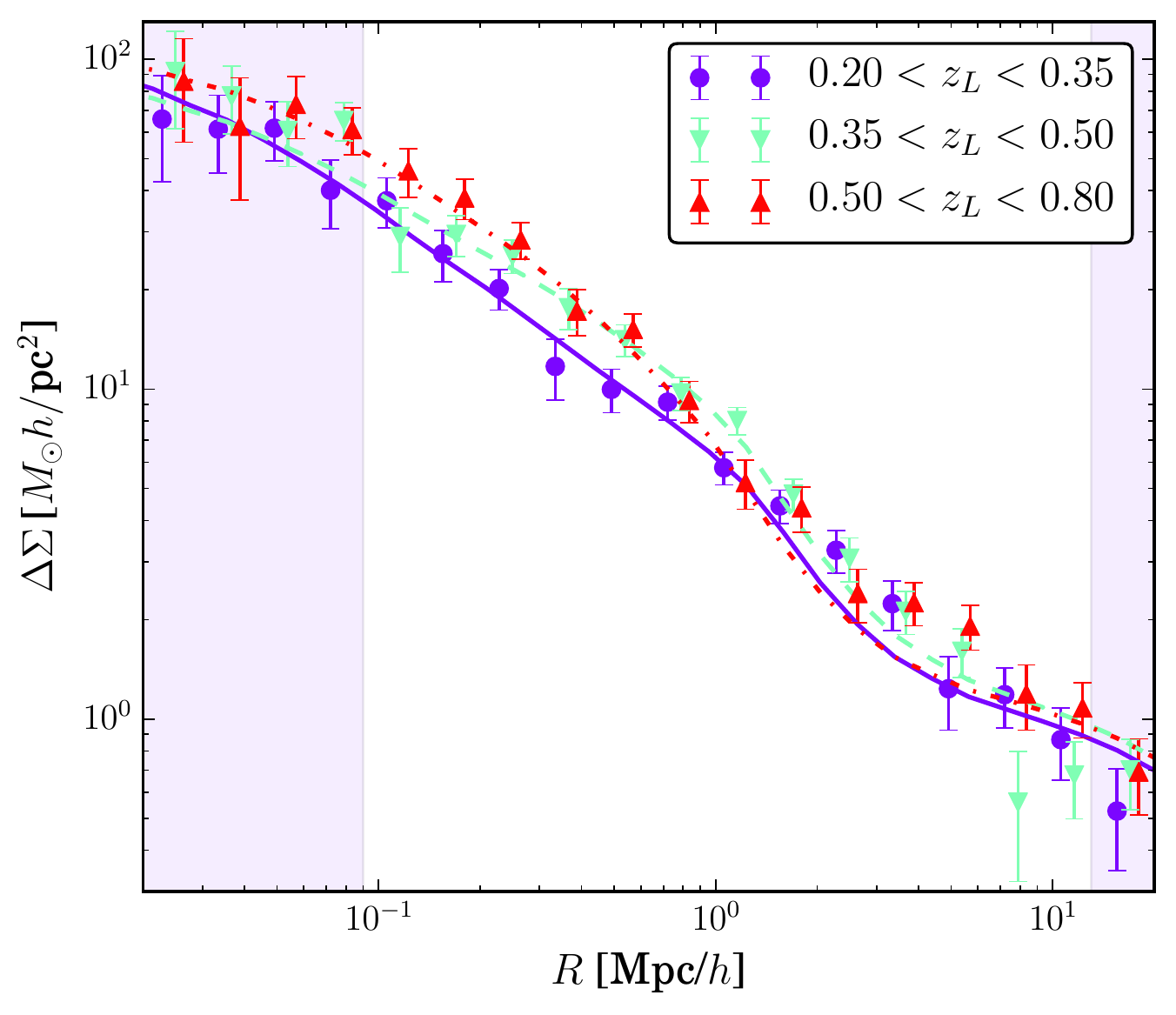}} \\
\resizebox{85mm}{!}{\includegraphics{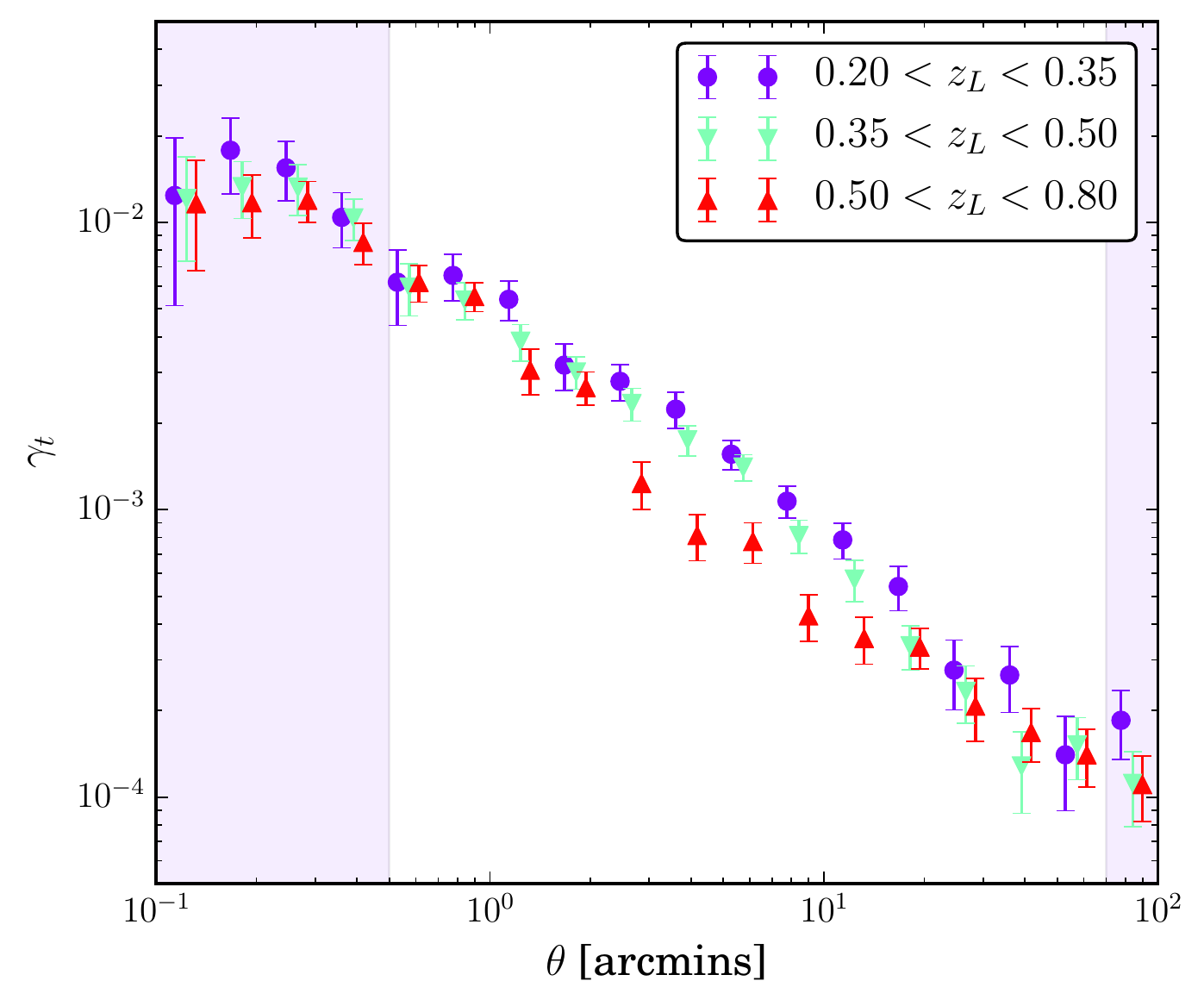}}
\caption{({\it upper panel}): $\Delta\Sigma$ measurement and statistical error bars for redMaGiC lenses in three redshift bins (as labelled).
Best-fit model curves are also shown for each sample.
The three different lens bins are consistent within our errors.
({\it lower panel}): The same, but showing the tangential shear $\gamma_t$.
}
\label{fig:deltas-meas}
\end{figure}
%%%%%%%%%%%%%%%%%%%

In Fig.~\ref{fig:deltas-meas} we show the measured signal and statistical errors for lens galaxies in three redshift bins.
The $\Delta\Sigma$ estimator uses the weighting in Eq.~(\ref{eq:dsig}), where the factor of $\Sigma_{\rm crit}^{-2}$ significantly downweights sources that are very close to the lens.
The three lens redshift samples are all consistent: taking the difference between any of the two samples and finding the $\chi^2$ with a null model (constant and zero) we find reduced $\chi^2 = 16.9/13, 19.7/13$, and $20.1/13$.
The latter two (both of which involve the middle redshift bin) are slightly high, but fitting a single parameter constant model to the difference between any pair of bins, we find consistent results.
The fractional difference between the pairs of bins is $0.12 \pm 0.2$, $-0.06 \pm 0.21$, and $0.17 \pm 0.13$:
the result is within $1\sigma$ of zero for two cases and just outside $1\sigma$ for the final case.
We conclude that our measurements are consistent with no evolution of the red sample, although they still leave open the possibility that future, higher signal-to-noise data will measure a difference.

Having checked that the measurements from every redshift bin are consistent, we also measure $\Delta\Sigma$ using the full redshift range, $0.2 < \zl < 0.8$.
We calculate the signal-to-noise of the measurement of the full sample as S/N $= \sqrt{\chi^2 - N_{\rm bin}} = 29$, where the $\chi^2$ is calculated using the jackknife covariance and a null model equal to zero.
We subtract the expectation value of the null hypothesis $\chi^2$ distribution, $N_{\rm bin}$, to make this an unbiased estimator of S/N.

We also show $\gamma_t$ measurements in Fig.~\ref{fig:deltas-meas} with the same lens samples, calculated according to Eq.~(\ref{eq:gam}).
For the lower and middle (upper) lens redshift bins, we use source redshifts $0.55 < \zs < 1.30$ ($0.83 < \zs < 1.30$).
The gap between lens and source redshifts is helpful in minimizing the inclusion of source galaxies that are actually in front of the lens, and therefore not lensed.
For $\gamma_t$ results with source tomography, and the implications for cosmology, see \citet{kwan17}.

%%%%%%%%%%%%%%%%%%%%%%%%%%%%
\subsection{Mean mass constraints}
\label{sec:mass}
%%%%%%%%%%%%%%%%%%%%%%%%%%%%

In this section, we use the measurements of $\Delta\Sigma(R)$ to
explore the dark matter environment of redMaGiC galaxies. We fit the
HOD model described in Section~\ref{sec:theory} with six free
parameters, M$_{\rm min}$, M$_1$, $\sigma$, $\alpha$, $f_{\rm cen}$, and $\sigma_8$.
We only consider scales between $0.09 < R < 15$
Mpc/$h$, due to deblending (Appendix~\ref{sec:deblending}) and
covariance (Appendix~\ref{sec:sims}) constraints.
We vary $\sigma_8$ along with HOD parameters because the mass is somewhat sensitive to $\sigma_8$ at large scales: $\sigma_8$ and bias are degenerate, the 2-halo term is proportional to bias, and bias is a monotonic function of mass in our model.
Our model fits are less sensitive to the other cosmology parameters, which we fix to
$\Omega_m = 0.31$, $h = 0.67$, $\Omega_b = 0.048$, $n_s = 0.96$, and $w = -1$,
all of which are consistent with the results of \citet{kwan17}.
Note that the results for redMaGiC galaxy bias are given in \citet{kwan17}, which uses large scale clustering in order to break the degeneracy between bias and $\sigma_8$.
We use the CosmoSIS package\footnote{\texttt{https://bitbucket.org/joezuntz/cosmosis}} \citep{zuntz15} to perform all fits.

\begin{figure}
\centering
\resizebox{85mm}{!}{\includegraphics{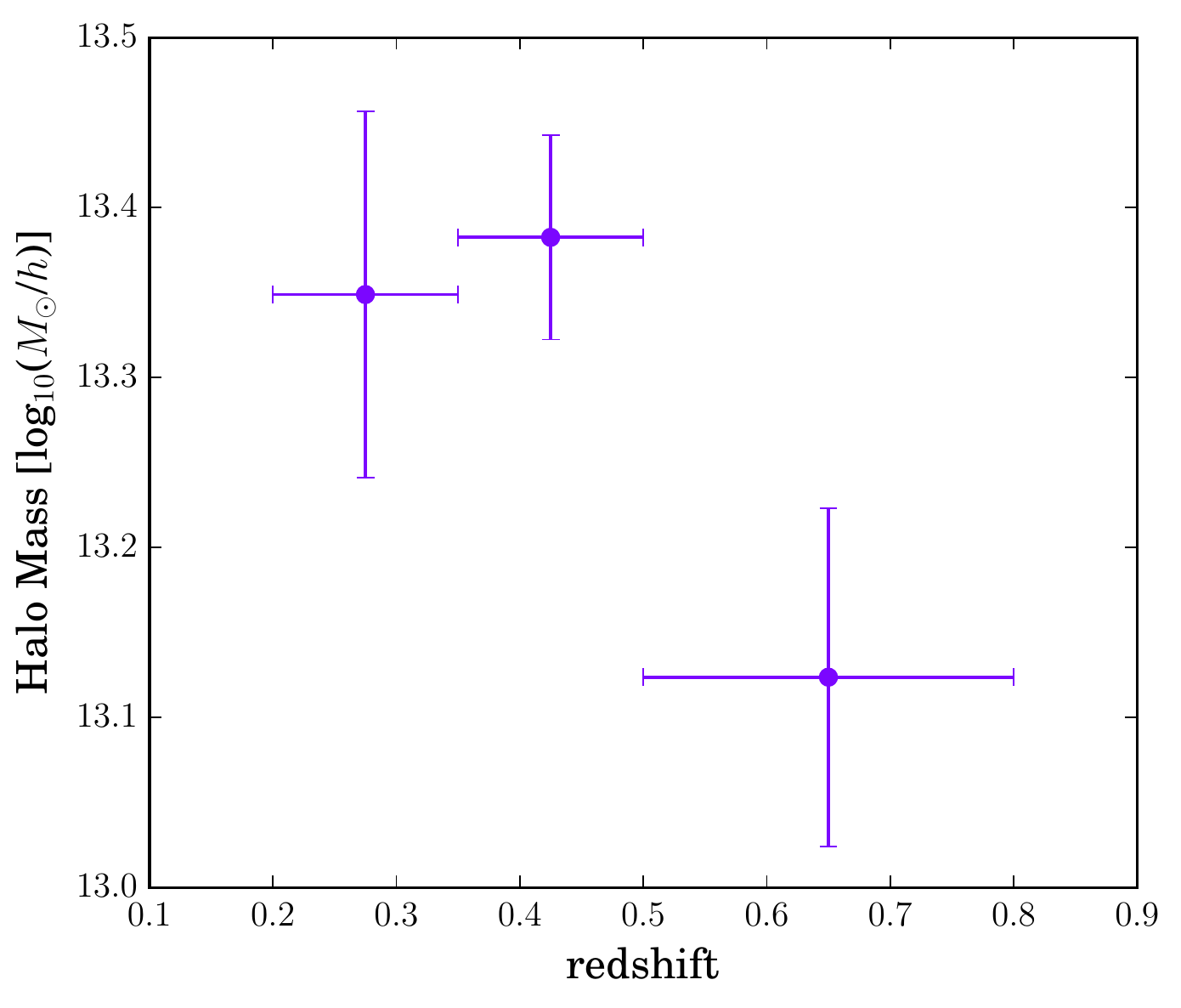}}
\caption{Redshift evolution of the mean halo mass of redMaGiC galaxies (purple points).
Taking into account pseudo-evolution and the covariance between lens bins, the mean mass is consistent with no evolution.
(See Sections~\ref{sec:measure} and \ref{sec:mass}.)}
\label{fig:hod_z_evol}
\end{figure}

Our best fit models are shown in Fig.~\ref{fig:deltas-meas} for each of the three lens bins.
The model goodness-of-fit is excellent in each case, with reduced $\chi^2 = 7.7 / 7, 10.6 / 7$, and $8.1 / 7$, in order of increasing redshift.
In Fig.~\ref{fig:hod_z_evol} we show constraints on the mean halo mass derived from the HOD
\begin{equation}
M_{\rm mean}=\frac{1}{\bar{n}}\int M_{\rm h} \frac{dn}{dM_{\mr{h}}}\ensav{N(M_{\mr h}|M_r^t)} dM_{\rm h} \, ,
\end{equation}
where $\frac{dn}{dM_{\mr{h}}}$ is the halo mass function and $\ensav{N(M_{\mr h}|M_r^t)}$ is the number of central and satellite galaxies.
The mean mass ranges from $\sim 10^{13.35} - 10^{13.12} M_\odot/h$ and shows little evolution between redshift bins.
Although the $z=0.5-0.8$ bin has a lower best-fit central value, much of the difference is due to pseudo-evolution \citep{diemer13}, the change in mass between halos at different redshifts due to defining mass relative to the mean matter density at that redshift.
Based on the results of \citet{diemer13}, this effect accounts for a drop in halo mass of $\Delta\log_{10}{M_{\rm mean}} \sim 0.1$ between $z = 0.3$ and 0.7.
This is roughly the size of our error bars, reducing the difference between bins to about $1\sigma$.
Note also that the errors in Fig.~\ref{fig:hod_z_evol} are correlated; our quantitative test for consistency of $\Delta\Sigma$ between different redshift bins (see Sec.~\ref{sec:measure}) takes into account these correlations.

Our results on $\sigma$ and $\alpha$ were strongly informed by our choice of priors.
$M_{\rm min}$ is constrained by our lensing data, but its central value is sensitive to the choice of priors in $\sigma$ and $\alpha$.
However, our key result for the mean mass of redMaGiC halos is not sensitive to the choice of priors.
We checked this by changing the priors significantly (e.g., doubling the prior width) from the fiducial choices in Table \ref{tab:parameters} and noting that the mean mass results of Fig.~\ref{fig:hod_z_evol} were unaffected.
Thus, at the level of our measurement errors, weak lensing is able to constrain the mass regardless of the uncertainty in the full HOD.

%%%%%%%%%%%%%%%%%%%
%%%%%%%%%%%%%%%%%%%

%%%%%%%%%%%%%%%
\section{Conclusions}
\label{sec:conclusions}
%%%%%%%%%%%%%%%

The main goal of this work was to validate galaxy-galaxy lensing measurements made with DES science verification data.
To that end we have performed a number of null tests on the shear catalogs and photometric redshifts and have quantified remaining systematic uncertainties.
We performed these null tests and all measurements with two independent shear pipelines, \texttt{ngmix} and \texttt{im3shape}, and found good consistency between the two methods.
The null tests and theory uncertainties are described in Sec.~\ref{sec:tests}.
One unresolved issue is the size split test with the \texttt{ngmix} catalog, which showed significant differences in $\Delta\Sigma$ measured from large and small source galaxies.
We discuss in Sec.~\ref{sec:split-size} the results of this test and why it is difficult to interpret, and leave its resolution for future work.
We validated our jackknife statistical errors using a suite of 50 mock surveys.
Such detailed tests are a necessary foundation for other work relying on tangential shear measurements with these data, for example, the cosmology results of \citet{kwan17} and \citet{baxter16}, bias results of \citet{prat16}, and trough \citep{gfa15} and void lensing results \citep{sanchez16}.

We measured the lensing signal of redMaGiC galaxies, a sample selected specifically to minimize photometric redshift error and outlier rate.
The total signal-to-noise of 29 allowed us to fit a simple HOD model and constrain the lens sample's central halo mass.
Dividing the lenses into 3 redshift bins over the range $0.2 < z < 0.8$, we found no evidence for evolution in the mean halo mass $\sim 2 \times 10^{13} M_\odot/h$ of our constant comoving density sample, at the level of current errors.

As the analysis begun here continues with DES Year 1 data and eventually the full 5 years of the survey, the statistical errors will continue to improve.
For example, assuming the full survey reaches the goal of 5000 square degrees with the same depth as the SV data, the volume probed by our lens sample will increase by a factor of 36.
Shape noise, our dominant small scale error, depends on the number of lenses as $1/\sqrt{N_{\rm lens}}$, for fixed source density.
Shape noise will thus be a factor of 6 smaller.
With this greater volume of data, new challenges will surface in ensuring the measurement is still statistics dominated.
This will require further work in understanding and modeling systematic uncertainties, but as those challenges are addressed our HOD constraints will improve quickly.
Another improvement for future work will involve adding information from galaxy clustering, as done by \citet{park16} in simulations.

%%%%%%%%%%%%%%%%%%%
\section*{Acknowledgments}

This paper has gone through internal review by the DES collaboration. We would also like to thank the external referee for helpful comments that improved the paper.

Funding for the DES Projects has been provided by the U.S. Department of Energy, the U.S. National Science 
Foundation, the Ministry of Science and Education of Spain, the Science and Technology Facilities Council of 
the United Kingdom, the Higher Education Funding Council for England, the National Center for Supercomputing 
Applications at the University of Illinois at Urbana-Champaign, the Kavli Institute of Cosmological Physics 
at the University of Chicago, the Center for Cosmology and Astro-Particle Physics at the Ohio State University, the Center for Particle Cosmology and the Warren Center at the University of Pennsylvania, 
the Mitchell Institute for Fundamental Physics and Astronomy at Texas A\&M University, Financiadora de 
Estudos e Projetos, Funda{\c c}{\~a}o Carlos Chagas Filho de Amparo {\`a} Pesquisa do Estado do Rio de 
Janeiro, Conselho Nacional de Desenvolvimento Cient{\'i}fico e Tecnol{\'o}gico and the Minist{\'e}rio da 
Ci{\^e}ncia e Tecnologia, the Deutsche Forschungsgemeinschaft and the Collaborating Institutions in the 
Dark Energy Survey. 

The Collaborating Institutions are Argonne National Laboratory, the University of California at Santa Cruz, 
the University of Cambridge, Centro de Investigaciones Energeticas, Medioambientales y Tecnologicas-Madrid, 
the University of Chicago, University College London, the DES-Brazil Consortium, the Eidgen{\"o}ssische 
Technische Hochschule (ETH) Z{\"u}rich, Fermi National Accelerator Laboratory,
the University of Edinburgh, 
the University of Illinois at Urbana-Champaign, the Institut de Ci\`encies de l'Espai (IEEC/CSIC), 
the Institut de F\'{\i}sica d'Altes Energies, Lawrence Berkeley National Laboratory, the Ludwig-Maximilians 
Universit{\"a}t and the associated Excellence Cluster Universe, the University of Michigan, the National Optical 
Astronomy Observatory, the University of Nottingham, The Ohio State University, the University of Pennsylvania, 
the University of Portsmouth, SLAC National Accelerator Laboratory, Stanford University, the University of 
Sussex, Texas A\&M University, and the OzDES Membership Consortium.

We are grateful for the extraordinary contributions of our CTIO colleagues and the DECam 
Construction, Commissioning and Science Verification teams in achieving the excellent 
instrument and telescope conditions that have made this work possible. The success of this 
project also relies critically on the expertise and dedication of the DES Data Management group.

The DES data management system is supported by the National Science Foundation under Grant Number 
AST-1138766. The DES participants from Spanish institutions are partially supported by MINECO under 
grants AYA2012-39559, ESP2013-48274, FPA2013-47986, and Centro de Excelencia Severo Ochoa 
SEV-2012-0234, some of which include ERDF funds from the European Union.

Support for DG was provided by NASA through the Einstein Fellowship Program, grant PF5-160138.

%%%%%%%%%%%%%%%%%%%
\appendix

%%%%%%%%%%%%%%%
\section{Systematics Tests}
\label{sec:syst-other}
%%%%%%%%%%%%%%%

Here we describe various systematic tests that were studied in other work, but are also relevant for our results.
These include deblending and sky subtraction, multiplicative shear bias, stellar contamination, and shear around stars.
The impact of these tests on our systematic uncertainty budget is summarized in Table~\ref{tab:syst}.

%%%%%%%%%%%%%%%%%%%%%%
\subsection{Deblending and sky subtraction}
\label{sec:deblending}
%%%%%%%%%%%%%%%%%%%%%%

Both shape measurement methods considered here fit parametric models across square ``postage-stamps'' of pixels centered on the galaxy being measured. These postage-stamps may contain light from neighboring objects that could bias the shape measurement. The direction of the bias is likely to be related to the direction of the neighbour with respect to the galaxy being measured - hence if the contaminating light is from the lens, or from objects spatially correlated with the lens, a small scale contamination to the tangential shear signal could arise. Any such effect is likely to be mitigated by the masking of neighbouring objects during fitting, and removing blended objects from the catalogs, as described in \citet{jsz15} (Sections 5.2 and 8.1 respectively). Figure 20 of \citet{jsz15} shows the tangential shear around bright stars as a function of angular separation, and this shows no evidence for a systematic signal around bright objects at small scales. Nonetheless we choose a conservative lower angular scale of 30 arcsec for the results in Sec.~\ref{sec:results}.

%%%%%%%%%%%%%%%%%%%%%%
\subsection{Multiplicative shear biases}
\label{sec:shear-bias}
%%%%%%%%%%%%%%%%%%%%%%

\citet{jsz15} studied in detail residual multiplicative biases for these shear catalogs.
They found that multiplicative bias should be less than 3\% in order to satisfy requirements for cosmic shear.
In Figure 24 of that work, they show that residual multiplicative biases are at most 1\% for \texttt{im3shape} and 2\% for \texttt{ngmix}.
The one exception is the lowest redshift bin for \texttt{ngmix}, which has a residual bias $\sim 4\%$.
However, since most of our signal comes from the higher redshift bins, we assume residual multiplicative bias is 2\%, and add this in quadrature with the other sources of error.

%%%%%%%%%%%%%%%%%%%%%%
\subsection{Stellar contamination and shear around stars}
\label{sec:stars}
%%%%%%%%%%%%%%%%%%%%%%

Since stars will not be gravitationally lensed by our lens galaxies, contamination of the source sample by stars will dilute our signal by the fraction of stars in the sample.
The DES SV galaxy clustering sample in \citet{ccb15} had at most 2\% stellar contamination.
While the sample selection for clustering differs somewhat from that for the weak lensing shear catalogs \citep{jsz15}, the differences should not increase stellar contamination.
Thus we take 2\% as our estimated systematic uncertainty from stellar contamination.

Similarly, stars do not act as gravitational lenses of distant source galaxies.
The measurement of tangential shear around faint stars provides a null test that can diagnose problems with PSF interpolation and PSF modelling \citep{jsz15}.
This measurement was shown for DES SV data in Figure 20 of \citet{jsz15}, and was consistent with the null hypothesis.

%%%%%%%%%%%%%%%
\section{Validation of statistical errors}
\label{sec:sims}
%%%%%%%%%%%%%%%

\begin{figure}
\centering
\resizebox{85mm}{!}{\includegraphics{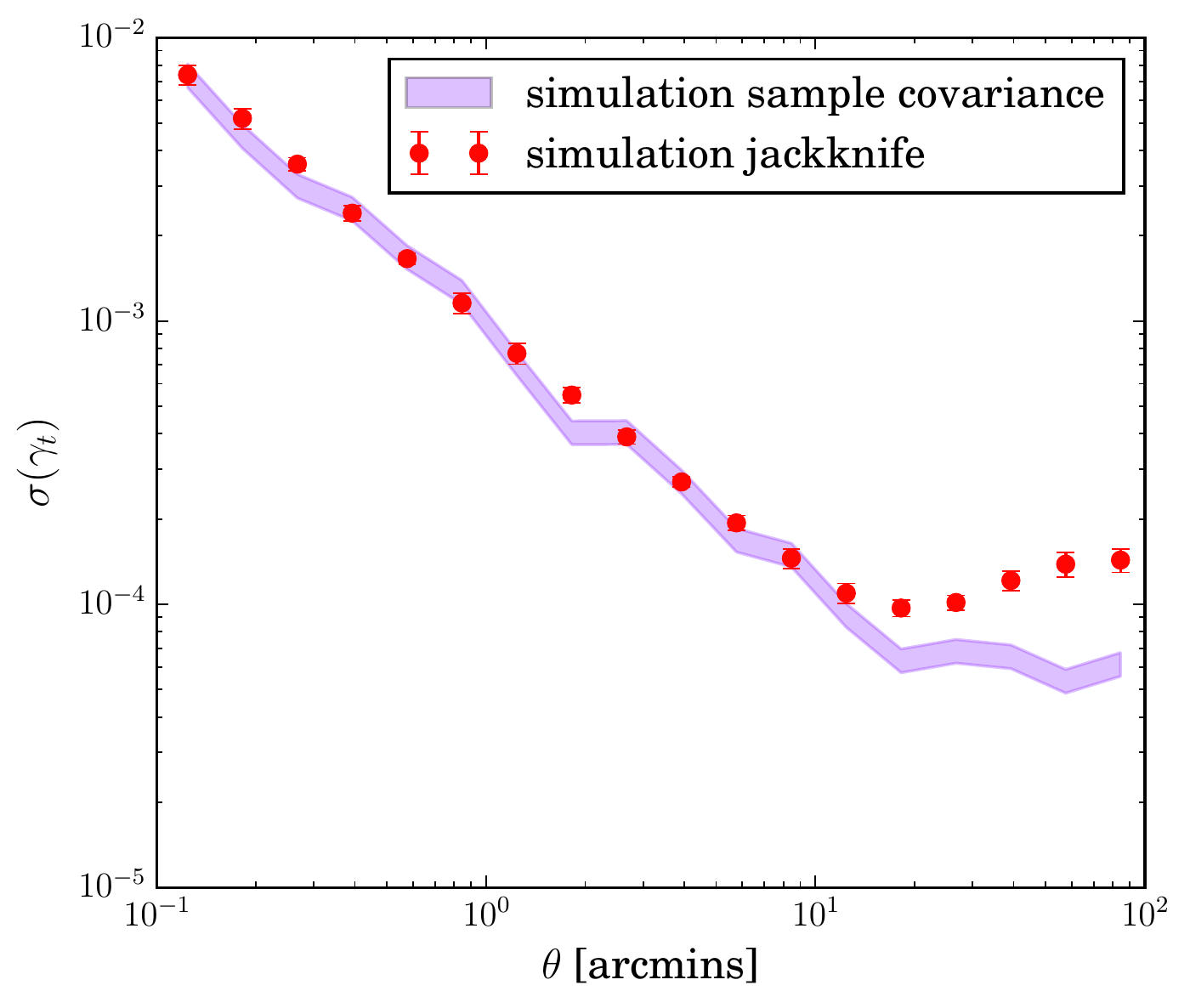}}
\caption{Comparison of diagonal errors from different methods: true covariance from independent simulations (purple band) and the jackknife method applied to the same simulations (red points).
The agreement between the true covariance and jackknife out to $\theta \sim 15$ arcmin validates the jackknife approach on these scales.
At $\theta \sim 70$ arcmin, our largest scale used in tests and fits, the jackknife method may overestimate the true error by a factor up to 2.
In this sense our HOD constraints using jackknife on the data are conservative.}
\label{fig:diag}
\end{figure}

\begin{figure}
\centering
\resizebox{41mm}{!}{\includegraphics{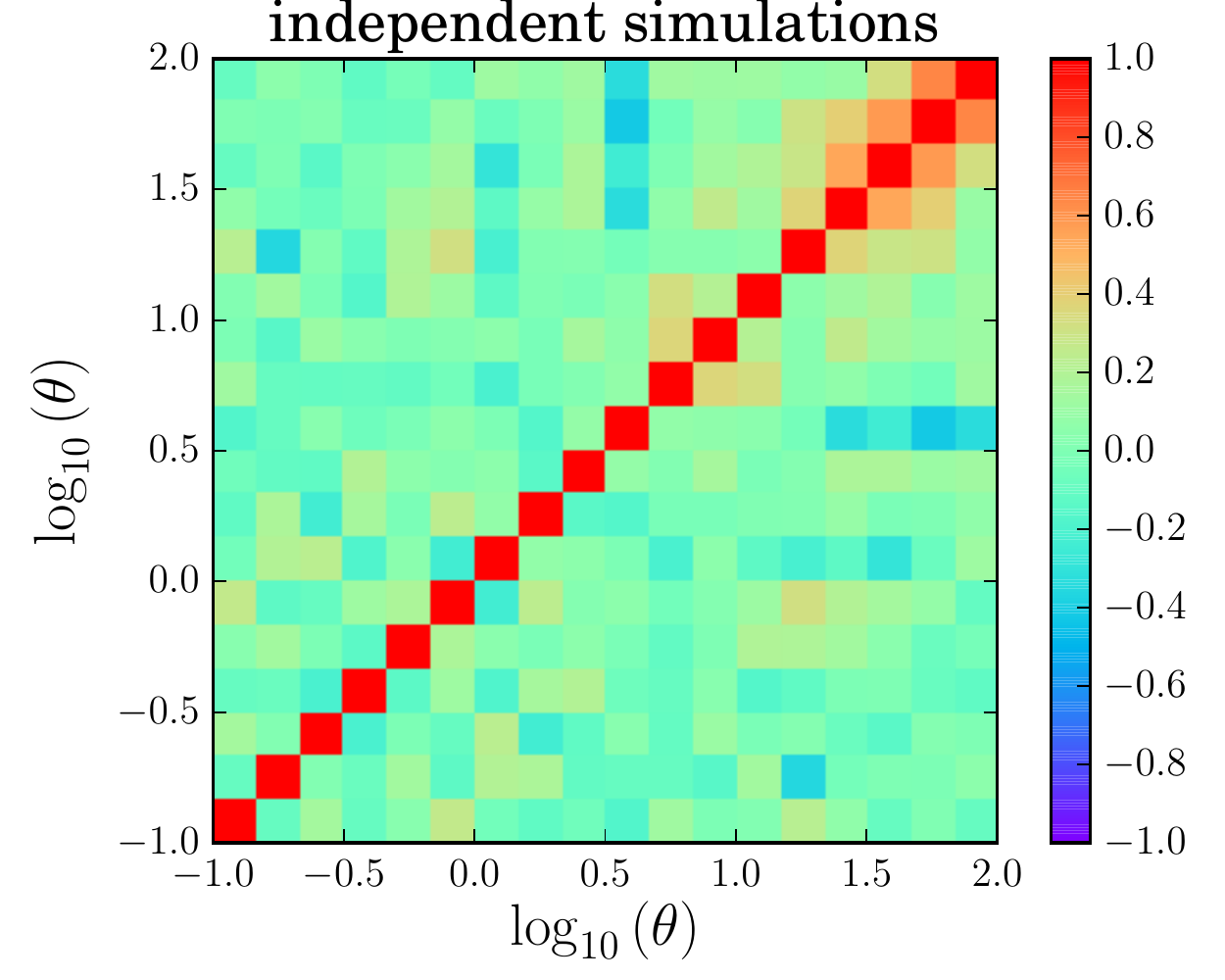}}
\resizebox{41mm}{!}{\includegraphics{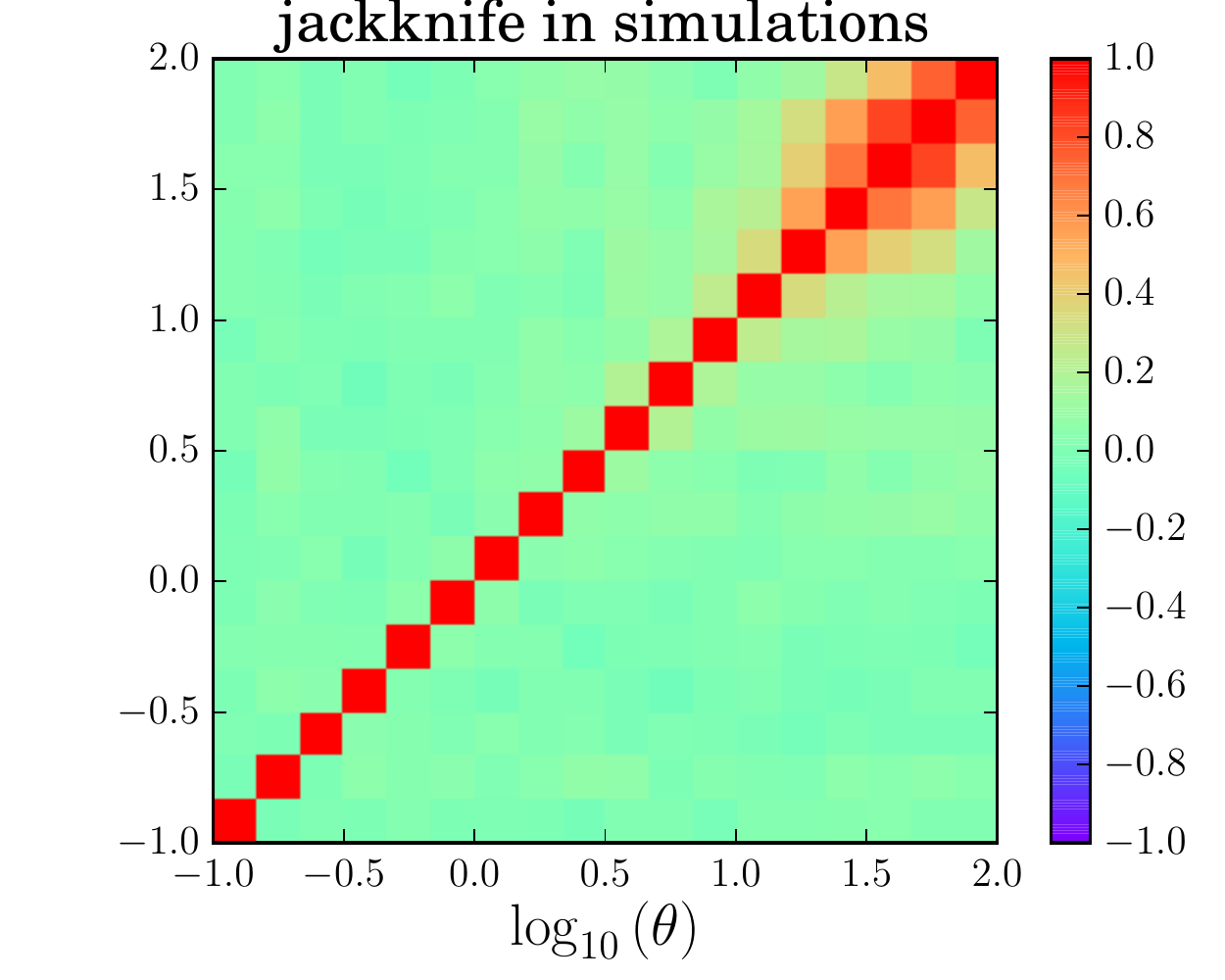}}
\caption{({\it left panel}): Normalized covariance matrix of $\gamma_t$ from 50 independent simulations, using lenses from $0.35 < \zl < 0.50$ and sources $0.83 < \zs < 1.3$.
({\it right panel}): The same, but showing the covariance from applying the jackknife method to our low redshift lens sample.
This covariance is less noisy since it is an average of the jackknife method applied to 10 simulations, but the qualitative features are very similar.
With both methods correlations are only significant on large scales, $\sim 10$ arcmin and above.
}
\label{fig:jk}
\end{figure}

In order to test the jackknife error bars obtained from the data, we compare to covariances from simulations.
The same redMaGiC algorithm has been run on mock galaxies in 50 nearly independent realizations of a 150 deg$^2$ survey constructed by dividing a wide area simulation into 50 pieces (see Sec.~\ref{sec:mock-catalog}). First, we compare the covariance from these independent realizations to errors obtained by dividing each 150 deg$^2$ simulation into 144 jackknife regions.
This comparison is made using the fiducial lens and source bins from \citet{kwan17}: lenses between $0.35 < \zl < 0.5$ and sources $0.83 < \zs < 1.30$.
The results are shown in Fig.~\ref{fig:diag}. The agreement is very good, at least out to $\sim 15$ arcminutes, where the jacknife method begins to systematically overestimate the true error.
Although this simulated area is slightly larger than our final area in the data and the number of realizations is smaller, the main point of this exercise is to validate the jackknife method.
Similarly, in Fig.~\ref{fig:jk} we compare the normalized covariances obtained with both methods.
The jackknife covariance is less noisy since it is an average of the jackknife method applied to 10 simulations, but the qualitative features are very similar.
Both methods have significant correlations just off the diagonal, starting around $\sim 10$ arcminutes.
At $\theta \sim 70$ arcmin, our largest scale used in tests and fits, the jackknife method may overestimate the true error by a factor up to 2.

%%%%%%%%%%%%%%%%%%%
%%%%%%%%%%%%%%%%%%%

%%%%%%%%%%%%%%%%%%%
%%%%%%%%%%%%%%%%%%%

%%%%%%%%%%%%%%%%%%%
% Bibliography
%%%%%%%%%%%%%%%%%%%

%%%%%%%%%%%%%%%%%%%

%%%%%%%%%%%%%%%%%%%
\section*{Affiliations}

$^1$ Department of Physics and Astronomy, University of Pennsylvania, Philadelphia, PA 19104, USA\\
$^2$ Institut de F\'{\i}sica d'Altes Energies (IFAE), The Barcelona Institute of Science and Technology, Campus UAB, 08193 Bellaterra (Barcelona) Spain\\
$^3$ Kavli Institute for Particle Astrophysics \& Cosmology, P. O. Box 2450, Stanford University, Stanford, CA 94305, USA\\
$^4$ Jodrell Bank Center for Astrophysics, School of Physics and Astronomy, University of Manchester, Oxford Road, Manchester, M13 9PL, UK\\
$^5$ Department of Physics, University of Arizona, Tucson, AZ 85721, USA\\
$^6$ Kavli Institute for Cosmological Physics, University of Chicago, Chicago, IL 60637, USA\\
$^7$ Department of Physics, University of Arizona, Tucson, AZ 85721, USA\\
$^8$ SLAC National Accelerator Laboratory, Menlo Park, CA 94025, USA\\
$^9$ Department of Physics, Stanford University, 382 Via Pueblo Mall, Stanford, CA 94305, USA\\
$^{10}$ Center for Cosmology and Astro-Particle Physics, The Ohio State University, Columbus, OH 43210, USA\\
$^{11}$ Institut de Ci\`encies de l'Espai, IEEC-CSIC, Campus UAB, Carrer de Can Magrans, s/n,  08193 Bellaterra, Barcelona, Spain\\
$^{12}$ Instituci\'o Catalana de Recerca i Estudis Avan\c{c}ats, E-08010 Barcelona, Spain\\
$^{13}$ Brookhaven National Laboratory, Bldg 510, Upton, NY 11973, USA\\
$^{14}$ Cerro Tololo Inter-American Observatory, National Optical Astronomy Observatory, Casilla 603, La Serena, Chile\\
$^{15}$ Department of Physics \& Astronomy, University College London, Gower Street, London, WC1E 6BT, UK\\
$^{16}$ Department of Physics and Electronics, Rhodes University, PO Box 94, Grahamstown, 6140, South Africa\\
$^{17}$ Department of Astrophysical Sciences, Princeton University, Peyton Hall, Princeton, NJ 08544, USA\\
$^{18}$ CNRS, UMR 7095, Institut d'Astrophysique de Paris, F-75014, Paris, France\\
$^{19}$ Sorbonne Universit\'es, UPMC Univ Paris 06, UMR 7095, Institut d'Astrophysique de Paris, F-75014, Paris, France\\
$^{20}$ Laborat\'orio Interinstitucional de e-Astronomia - LIneA, Rua Gal. Jos\'e Cristino 77, Rio de Janeiro, RJ - 20921-400, Brazil\\
$^{21}$ Observat\'orio Nacional, Rua Gal. Jos\'e Cristino 77, Rio de Janeiro, RJ - 20921-400, Brazil\\
$^{22}$ Department of Astronomy, University of Illinois, 1002 W. Green Street, Urbana, IL 61801, USA\\
$^{23}$ National Center for Supercomputing Applications, 1205 West Clark St., Urbana, IL 61801, USA\\
$^{24}$ Institute of Cosmology \& Gravitation, University of Portsmouth, Portsmouth, PO1 3FX, UK\\
$^{25}$ School of Physics and Astronomy, University of Southampton,  Southampton, SO17 1BJ, UK\\
$^{26}$ Faculty of Physics, Ludwig-Maximilians-Universit\""at, Scheinerstr. 1, 81679 Munich, Germany\\
$^{27}$ Excellence Cluster Universe, Boltzmannstr.\ 2, 85748 Garching, Germany\\
$^{28}$ Fermi National Accelerator Laboratory, P. O. Box 500, Batavia, IL 60510, USA\\
$^{29}$ Department of Astronomy, University of Michigan, Ann Arbor, MI 48109, USA\\
$^{30}$ Department of Physics, University of Michigan, Ann Arbor, MI 48109, USA\\
$^{31}$ Department of Physics, The Ohio State University, Columbus, OH 43210, USA\\
$^{32}$ Australian Astronomical Observatory, North Ryde, NSW 2113, Australia\\
$^{32}$ Departamento de F\'{\i}sica Matem\'atica,  Instituto de F\'{\i}sica, Universidade de S\~ao Paulo,  CP 66318, CEP 05314-970, S\~ao Paulo, SP,  Brazil\\
$^{33}$ Laborat\'orio Interinstitucional de e-Astronomia - LIneA, Rua Gal. Jos\'e Cristino 77, Rio de Janeiro, RJ - 20921-400, Brazil\\
$^{34}$ George P. and Cynthia Woods Mitchell Institute for Fundamental Physics and Astronomy, and Department of Physics and Astronomy, Texas A\&M University, College Station, TX 77843,  USA\\
$^{35}$ Department of Astronomy, The Ohio State University, Columbus, OH 43210, USA\\
$^{36}$ Max Planck Institute for Extraterrestrial Physics, Giessenbachstrasse, 85748 Garching, Germany\\
$^{37}$ Jet Propulsion Laboratory, California Institute of Technology, 4800 Oak Grove Dr., Pasadena, CA 91109, USA\\
$^{38}$ Department of Physics and Astronomy, Pevensey Building, University of Sussex, Brighton, BN1 9QH, UK\\
$^{39}$ Centro de Investigaciones Energ\'eticas, Medioambientales y Tecnol\'ogicas (CIEMAT), Madrid, Spain\\
$^{40}$ Computer Science and Mathematics Division, Oak Ridge National Laboratory, Oak Ridge, TN 37831\\
$^{41}$ Argonne National Laboratory, 9700 South Cass Avenue, Lemont, IL 60439, USA\\
$\dagger$ Einstein Fellow

\end{document}